\documentclass[acmtocl]{acmtrans2m}
\usepackage{amssymb}
 \usepackage{amsmath}

\newdef{definition}[theorem]{Definition}
\newdef{remark}[theorem]{Remark}
\usepackage{fullpage}
 \usepackage{graphicx}
 
 \usepackage{placeins} % for floatbarrier
\usepackage{nicefrac}
\usepackage{setspace}
 
\begin{document}

\markboth{M. Kumar \& A. M. Mathai}{Design of variable acceptance sampling plan for exponential distribution under uncertainty}

\title{Design of Variable Acceptance Sampling Plan for Exponential Distribution under Uncertainty}
            
\author{M. KUMAR and A. M. MATHAI \\ National Institute of Technology Calicut}

\begin{abstract} 
 In an acceptance monitoring system, acceptance sampling techniques are used to increase production, enhance control, and deliver higher-quality products at a lesser cost. It might not always be possible to define the acceptance sampling plan parameters as exact values, especially, when data has uncertainty.
In this work, acceptance sampling plans for a large number of identical units with exponential lifetimes are obtained by treating acceptable quality life, rejectable quality life, consumer's risk, and producer's risk as fuzzy parameters. To obtain plan parameters of sequential sampling plans and repetitive group sampling plans, fuzzy hypothesis test is considered. To validate the sampling plans obtained in this work, some examples are presented. Our results are compared with existing results in the literature. Finally, to demonstrate the application of the resulting sampling plans, a real-life case study is presented        
\end{abstract}
            
\category{62K05}{Statistics}{Design of statistical experiments}[Optimal statistical designs]

\category{62K25}{Statistics}{Design of statistical experiments}[Fuzziness and design of statistical experiments]

\category{62K99}{Statistics}{Design of statistical experiments}

\category{62P30}{Statistics}{Applications of statistics}[Applications of statistics in engineering and industry; control charts]
            
\terms{Design; Reliability.} 
            
\keywords{ Exponential distribution, fuzzy acceptance sampling plan, fuzzy hypothesis testing, fuzzy optimization.}

\begin{bottomstuff} 
Author's address: M. Kumar and A. M. Mathai, Department of Mathematics, National Institute of Technology, Calicut, Kerala, India, PIN - 673601.\\
E-mail address: mahesh@nitc.ac.in (M. Kumar) (Corresponding author) and ashlyn\_p190073ma@nitc.ac.in (A. M. Mathai).
\end{bottomstuff}
         
\maketitle

\section{Introduction}
Acceptance sampling plan (ASP) for life testing has a unique role in the field of reliability and quality control. In the field of manufacturing, there is a need for reliable plan which is  economical in terms of time and manufacturing costs. The classical sampling plans mainly depend upon the accuracy of data, the test of hypotheses, and appropriate decision rules. At the same time, existing literature has shown the existence of situations where the above assumptions are rather unrealistic or vague (see \cite{watanabe1993,sadeghpour2011acceptance}). The uncertainties of parameters in the sampling plans are due to season, trend, and random variations. In such cases of sampling inspection, the data will be recorded as linguistic variables, and the fuzzy set theory proposed by \citeN{klir1996fuzzy} is used to analyze uncertain data.
\citeN{sherman1965} explained an acceptance sampling plan for repetitive group sampling, in which the decision of accepting and rejecting the lot depends on the number of defectives. Probabilities for long-run acceptance, rejection, and the number of samples to get an acceptance decision are also formulated. Different methods have been considered by many authors in the literature to design acceptance sampling plans and most of them are concentrating on the reduction of sample size but not the cost (see  \cite{sherman1965,aslam2012decision}). \citeN{MK} designed a new optimal acceptance sampling plan by variables for a lot of units having exponentially, distributed lifetime based on the time between successive failures. They considered both the methods of sequential sampling plan and repetitive group sampling plans under different criteria. In all these, an optimization problem is performed to minimize the expected testing cost under Type I and Type II errors to obtain the optimal parameters.

All the above sampling plans in the literature assume that data do not have uncertainties. In other words, these sampling plans are designed for crisp data, which are not practical when real-life problems are concerned where uncertainties in data are unavoidable. 
\citeN{torabi2009} redefined the concepts of fuzzy hypotheses testing by defining the weighted probability distribution function. They introduced the sequential probability ratio test (SPRT) by attributes for fuzzy hypotheses testing and illustrated several examples.
\citeN{jameel2012} had proposed a sequential sampling plan by variables using SPRT for fuzzy hypothesis in the case of a lot of units having a normally distributed lifetime. According to their plan, the width of the continuation of testing in the graph of the sampling plan narrowed as the sample size increased.

Fuzzy hypothesis testing was studied via various methods by several authors [\citeNP{delgado1985}, \citeNP{watanabe1993}, \citeNP{arnold1995},  \citeyearNP{arnold1996},  \citeNP{taheri2001}, \citeNP{torabi2007likelihood}, \citeNP{jamkhaneh2010}]. \citeN{jameel2012}, \citeN{sakawa2013} and \citeN{trappey1988} discussed the  optimization of fuzzy non-linear programming problem (NLPP) by converting it to crisp NLPP in the cases of fuzzy constraints.

 In previous literature, fuzzy hypothesis testing is done using SPRT and the likelihood ratio test. The decision for accepting and rejecting the lot is made by the cumulative averages. The optimization was concentrated on sample size and failure time. In this paper, we consider a lot of identical units having exponential lifetime distribution. Here we have obtained sampling plans based on the time between successive failures, which is so far not considered in the literature. It is noted that the design of ASP by considering the time between successive failures has an advantage in bringing down the test cost involved in ASP (for example see \citeN{MK}). Further, the obtained ASP guarantees minimum testing cost which is novel in the design of fuzzy ASP. 
 
 This paper is organized as follows: In Section 2, the sequential sampling plan is discussed with fuzzy parameters. A fuzzy optimization problem is constructed  and solved to minimize the testing cost subject to fuzzy constraints of Type I and II errors. In Section 3, three different methods of repetitive group sampling for fuzzy hypothesis testing are considered. In the first method, the lot is accepted or rejected by checking the minimum of \textit{n} samples under the fuzzy hypothesis. In the second criterion, we consider the maximum of \textit{n} samples to make a decision. Finally, an ASP is derived using data obtained from Type I censoring with the fuzzy hypothesis is discussed. The fuzzy optimization problem to minimize the testing cost subject to fuzzy Type I and Type II errors is solved to get optimal parameters of sampling plans in all three methods. Moreover, in all these cases, a comparative study is done with the existing crisp methods. In Section 4, the de-fuzzification of all fuzzy parameters using the center of gravity method is discussed and it is observed that the weighted probability distribution method is better than de-fuzzification. Section 5 depicts a comparison of all sampling plans using various numerical examples and a real-life case study.
\section{Sequential Sampling Plan with Fuzzy Parameters}
In this section, we define a sequential sampling plan (SSP), for fuzzy hypothesis and  for that consider a lot of units with exponential failure time with probability density function given by
 \begin{equation}
f(x)= \begin{cases}\lambda \exp \left(-\lambda x\right), & x>0, \lambda>0, \\ 0, & \text { otherwise. }\end{cases} 
\label{2.1}
\end{equation}
Let $X_{i}$ denote the  $i^{th}$ failure time, $i \in \mathbb{N}$. The  acceptance sampling plan is based on the time between two consecutive failures (see \citeN{MK}) denoted by $Y_{i+1}=X_{i+1}-X_{i}, \ i=1,2, \ldots$ and $Y_{1}=X_{1}$. Also, $Y_{i}$, $i=1,2,3, \ldots$ follows exponential distribution with same parameter $\lambda$. The following is the ASP:
At any point of failure,
 \begin{center}
  continue the process, if $\quad t_{1} \leq Y_{i}<t_{2}$,\\
accept the lot, if $\quad Y_{i} \geq t_{2}$,\\
reject the lot, if $\quad Y_{i}<t_{1}.$  
 \end{center} 
   The following are the hypotheses for SSP as a statistical test, for accepting or rejecting the lot:
$$ \left\{\begin{array}{l}
H_{0}: \text{The lot is of acceptable quality life (AQL)}, \\
H_{1}: \text{The lot is of rejectable quality life (RQL)}.
\end{array}\right. $$
The probability of Type I and II errors $\alpha$ and $\beta$ for the acceptance and rejection of the lot of SSP in crisp case is described in \citeN{MK} as follows:
\begin{equation}
    \begin{split}
     &P\left( \text{Reject the lot}  \mid \frac{1}{\lambda} \geq \lambda_0 \right) \leq \alpha,\\ 
      &P\left( \text{Accept the lot}  \mid \frac{1}{\lambda} \leq \lambda_1  \right) \leq \beta, 
      \label{2.2}
    \end{split}
\end{equation}
where $\alpha$ is the producer's risk and $\beta$ is the consumer's risk.
According to \citeN{wald2004}, by the definition of the ASP used here, the inequalities defined in~(\ref{2.2}) will be equivalent to the following
\begin{equation*}
    \begin{split}
    \qquad \qquad \qquad \qquad \quad  &P\left( \text{Reject the lot}  \mid \frac{1}{\lambda} = \lambda_0 \right) \leq \alpha,\\ 
      &P\left( \text{Accept the lot}  \mid \frac{1}{\lambda} = \lambda_1  \right) \leq \beta. \qquad \qquad \qquad \quad
    \end{split}
\end{equation*}
 In most real-life situations, while handling the data, the parameters, AQL, RQL, $\alpha$, and $\beta$ will be expressed as linguistic variables and they can be converted to fuzzy parameters.
Let $\lambda_{0}$ denote the acceptable quality life (AQL) and $\lambda_{1}$ the rejectable quality life (RQL) of an item in the lot.

\begin{figure}[!htb]
  \centering
 \includegraphics[width=7cm,height=5cm]{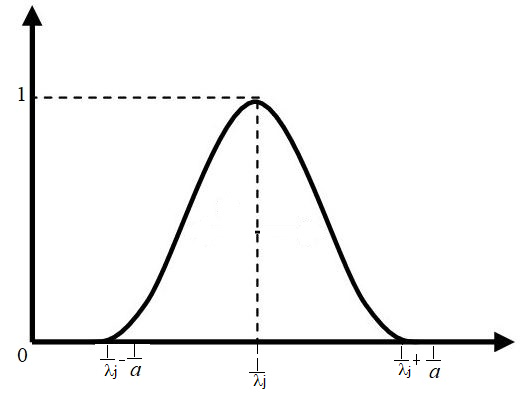}
  \caption{Membership function of $\frac{1}{\lambda_j}, \ j = 0, 1.$}
  \label{fig:Fig1}
\end{figure}

Let $\lambda_0$, $\lambda_1$, $\alpha$, and $\beta$ be fuzzy numbers. Then the fuzzy hypothesis is given by:
\begin{equation}
    \begin{split}
     H_{0}&: \text{average life is approximately AQL},\\
    H_{1}&: \text{average life is approximately RQL}. 
    \end{split}
    \label{2.3}
\end{equation}

$$ \left\{\begin{array}{l}
H_{0}: \frac{1}{\lambda} \approx \lambda_{0}, \\
H_{1}: \frac{1}{\lambda} \approx \lambda_{1}.
\end{array}\right.\qquad \quad$$
That is,
$$\left\{\begin{array}{l}
H_{0}: \lambda \approx \frac{1}{\lambda_{0}}, \\
H_{1}: \lambda \approx \frac{1}{\lambda_{1}},
\end{array}\right.\qquad \quad $$
$$ \left\{\begin{array}{l}
H_{0}: \lambda \quad \textrm{is} \quad H_{0}(\lambda), \\
H_{1}: \lambda \quad \textrm{is} \quad H_{1}(\lambda),
\end{array}\right.$$
where the membership function, $H_j(\lambda),\ j= 0, 1$ (see Figure~\ref{fig:Fig1}) is given by [\citeNP{palaniappan2005fuzzy}, \citeNP{olunloyo2009design}]. That is,
$$H_j(\lambda)=\left\{\begin{array}{ll}
\frac{1+cos(\textit{a}\pi(\lambda-\frac{1}{\lambda_j}))}{2}  & , \quad \frac{1}{\lambda_j} - \frac{1}{\textit{a}} \leq \lambda \leq \frac{1}{\lambda_j} + \frac{1}{\textit{a}}, \\
0  & ,\quad \text { otherwise, } \qquad \qquad \qquad j= 0, 1.
\end{array}\right.$$
$$a > \lambda_j , j = 0,1 \quad \textrm{and} \quad a > 0.$$

The following probability requirements will be used to determine whether to accept or reject the lot:
\begin{equation}
     P\left( \text{Reject the lot}  \mid \lambda \approx \frac{1}{\lambda_0}\right) \leq \alpha, 
     \label{2.4}
     \end{equation}
     \vspace{-5mm}
     \begin{equation}
P\left( \text{Accept the lot}  \mid \lambda \approx \frac{1}{\lambda_1}\right) \leq \beta.  \label{2.5}
\end{equation}

Our goal is to find an optimal pair ($t_1,t_2 $) such that the total cost (TC) of testing is minimized subject to the Type I and Type II errors given by~(\ref{2.4}) and~(\ref{2.5}) when all the parameters, namely, AQL, RQL, $\alpha$, and $\beta$ are fuzzy.

The overall cost of testing is calculated by multiplying the testing cost of an item per unit time by the entire duration of testing.
The failure time of an item is the random variable $Y_i$, then the duration of the test is $Y_i$ times the number of failed items. Note that, the overall cost of testing (TC) is a random variable.

Let $p_a$, $p_r$ and $p_c$ denote the probability of acceptance, rejection of the lot and continuation of testing respectively, while the failure of an item is observed. The overall probability of acceptance, rejection are determined by $P_{A}=\frac{P_{a}}{1-p_{c}},   P_{R}=\frac{p_{r}}{1-p_{c}}$ and expected number of failed items by $N=\frac{1}{1-p_{c}}$ (see \citeN{sherman1965}).

Under the hypothesis $H_j, j=0,1$ the weighted probability density function (WPDF) (see \citeN{torabi2009}) of $Y_i$ is given by,
\begin{align*}
\begin{split}
     f_{j}(y)=\biggr\{\left[\left(\textit{\textrm{a}}-\lambda_{j}\right) y^{3}+3 \textit{\textrm{a}} y^{2} \lambda_{j}+\pi^{2} \textit{\textrm{a}}^{2}\left(\textit{\textrm{a}}-\lambda_{j}\right) y +\pi^{2} \textit{\textrm{a}}^{3} \lambda_{j}\right] e^{\frac{2 y}{\textit{\textrm{a}}}}- (\lambda_j + \textrm{\textit{a}}) y^{3}\\
             -3 \textit{\textrm{a}} y^{2} \lambda_{j}-\pi^{2} \textit{\textrm{a}}^{2}(\textrm{\textit{a}}+\lambda_j)y - \pi^{2} \textit{\textrm{a}}^{3} \lambda_{j}\biggr\} \frac{\pi^{2} \textit{\textrm{a}}^{2} e^{-\frac{y}{\textit{\textrm{a}}}-\frac{y}{\lambda}_{j}}}{2y^2 \left(y^2 + \pi^2 \textit{\textrm{a}} \right)^2 \lambda_j}.
     \end{split}
\end{align*}
\begin{equation}
 p_{c}=P\left(t_{1} \leq Y_{i}<t_{2}\right) = \frac{\pi^2 \textit{a}^3 }{2}\left[ \frac{\left(e^{\frac{2t_{1}}{\textit{a}}}-1\right)e^{-\frac{t_{1}}{\lambda_j}-\frac{t_{1}}{\textit{a}}}}{t_{1}^3 + \pi^2 \textit{a}^2 t_{1} }- \frac{\left(e^{\frac{2t_{2}}{\textit{a}}}-1\right)e^{-\frac{t_{2}}{\lambda_j}-\frac{t_{2}}{\textit{a}}}}{t_{2}^3 + \pi^2 \textit{a}^2 t_{2} }\right], 
 \label{2.6}
 \end{equation}
 \begin{equation}
p_{\textit{\textit{a}}}=P\left(Y_{i} \geq t_{2}\right) = \frac{\pi^2 \textit{a}^3 \left(e^{\frac{2t_{2}}{\textit{a}}}-1\right)e^{-\frac{t_{2}}{\lambda_j}-\frac{t_{2}}{\textit{a}}}}{2t_{2}^3 + 2\pi^2 \textit{a}^2 t_{2}},
\label{2.7}
\end{equation}
\begin{equation}
p_{r}=P\left(Y_{i}<t_{1}\right) = 1 - \frac{\pi^2 \textit{a}^3 \left(e^{\frac{2t_{1}}{\textit{a}}}-1\right)e^{-\frac{t_{1}}{\lambda_j}-\frac{t_{1}}{\textit{a}}}}{2t_{1}^3 + 2\pi^2 \textit{a}^2 t_{1}}. 
\label{2.8}
\end{equation}
\subsection{Model development and fuzzy optimization problem}
Our aim is to minimize TC such that conditions~(\ref{2.4}) and~(\ref{2.5}) are satisfied. We note that the optimization problem with a random objective for TC leads to an intractable problem, and cannot be solved analytically by satisfying the constraints in~(\ref{2.4}) and~(\ref{2.5}). Hence, as in \citeN{MK}, we consider a problem of minimizing the expected testing cost (ETC) satisfying~(\ref{2.4}) and~(\ref{2.5}). Considering $C$ as the cost for testing an item for unit time and $E_0(Y_i)$ as the expectation of $Y_i$ under $H_0$, we have
\begin{equation*}
 \mathrm{Total \ ETC}=\frac{C \times E_0(Y_i)}{1-p_{c}},   
\end{equation*}
 where $E_0(Y_i)$ is given by,
\begin{equation*}
   E_0(Y_i) = \frac{\textrm{\textit{a}}}{2}  \operatorname{ln} \left[\frac{\textrm{\textit{a}}+\lambda_0}{\textrm{\textit{a}}-\lambda_0}\right]+ \frac{\textrm{\textit{a}}}{2}\left[
\cos \left( \frac{\textrm{\textit{a}}\pi}{\lambda_0}\right) \displaystyle\int\limits_{ \frac{\textrm{\textit{a}}\pi}{\lambda_0}-\pi}^{ \frac{\textrm{\textit{a}}\pi}{\lambda_0}+\pi} \frac{\cos u}{u} d u+\sin \left(\frac{\textrm{\textit{a}}\pi}{\lambda_0} \right) \displaystyle\int\limits_{\frac{\textrm{\textit{a}}\pi}{\lambda_0} -\pi}^{\frac{\textrm{\textit{a}}\pi}{\lambda_0}+\pi} \frac{\sin u}{u} d u\right].  
\end{equation*}
 The inequality in the constraints~(\ref{2.4}) and~(\ref{2.5}) will be more relaxed than the crisp case and the tilda sign denotes the fuzziness in the parameters $\alpha$ and $\beta$.  According to ASP, $\lambda_0 \geq t_2$ as it is the AQL. Thus, the optimization problem under fuzziness in parameters is formulated as,
\begin{equation} 
\begin{aligned}[b]
P_1(t_1, t_2):\qquad &\underset{\left(t_{1}, t_{2}\right)}{\operatorname{Min}} \left(\frac{C E_0(Y_i)}{1-p_{c}}\right) \text {subject to} \\
&P\left( \text{Reject the lot}  \mid \lambda \approx \frac{1}{\lambda_0}\right) \stackrel{\sim}{\leq} \alpha, \label{2.9}  \end{aligned}
\end{equation}
\begin{equation} 
\qquad \qquad \quad \qquad P\left(\text {Accept the lot} \mid \lambda \approx \frac{1}{\lambda_{1}}\right) \stackrel{\sim}{\leq} \beta, \label{2.10}
\end{equation}
where $t_{1}, t_{2}>0$ and $t_{2}>t_{1}$. That is,
\begin{equation*}
\begin{split}
\quad&\underset{\left(t_{1}, t_{2}\right)}{\operatorname{Min}}\left(\frac{C E_0(Y_i)}{1-p_{c}}\right) \text { subject to} \\
&\left(\frac{p_{r}}{1-p_{c}} \mid \lambda \approx \frac{1}{\lambda_{0}}\right) \stackrel{\sim}{\leq} \alpha, \\
&\left(\frac{p_{a}}{1-p_{c}} \mid \lambda \approx \frac{1}{\lambda_{1}}\right) \stackrel{\sim}{\leq} \beta,
\end{split}
\end{equation*}
where $t_{1}, t_{2}>0$ and $t_{2}>t_{1}$. Substituting for $p_{c}, p_{a}$ and $p_{r}$ from~(\ref{2.6}),~(\ref{2.7}) and~(\ref{2.8}), the fuzzy optimization problem, $P_1(t_1, t_2)$, becomes problem $P_1'(t_1, t_2)$:
$$P_1'(t_1, t_2): \qquad \underset{\left(t_{1}, t_{2}\right)}{\operatorname{Min}}\quad  \frac{C E_0(Y_i)}{1- \frac{\pi^2 \textit{a}^3 }{2}\left[ \frac{\left(e^{\frac{2t_{1}}{\textit{a}}}-1\right)e^{-\frac{t_{1}}{\lambda_0}-\frac{t_{1}}{\textit{a}}}}{t_{1}^3 + \pi^2 \textit{a}^2 t_{1} }- \frac{\left(e^{\frac{2t_{2}}{\textit{a}}}-1\right)e^{-\frac{t_{2}}{\lambda_0}-\frac{t_{2}}{\textit{a}}}}{t_{2}^3 + \pi^2 \textit{a}^2 t_{2} }\right]}$$
\begin{equation}
\begin{aligned}[b]
\text { such that} \quad &\frac{1 - \frac{\pi^2 \textit{a}^3 \left(e^{\frac{2t_{1}}{\textit{a}}}-1\right)e^{-\frac{t_{1}}{\lambda_0}-\frac{t_{1}}{\textit{a}}}}{2t_{1}^3 + 2\pi^2 \textit{a}^2 t_{1}}}{1- \frac{\pi^2 \textit{a}^3 }{2}\left[ \frac{\left(e^{\frac{2t_{1}}{\textit{a}}}-1\right)e^{-\frac{t_{1}}{\lambda_0}-\frac{t_{1}}{\textit{a}}}}{t_{1}^3 + \pi^2 \textit{a}^2 t_{1} }- \frac{\left(e^{\frac{2t_{2}}{\textit{a}}}-1\right)e^{-\frac{t_{2}}{\lambda_0}-\frac{t_{2}}{\textit{a}}}}{t_{2}^3 + \pi^2 \textit{a}^2 t_{2} }\right]} \stackrel{\sim}{\leq} \alpha,\\
&\frac{\frac{\pi^2 \textit{a}^3 \left(e^{\frac{2t_{2}}{\textit{a}}}-1\right)e^{-\frac{t_{2}}{\lambda_1}-\frac{t_{2}}{\textit{a}}}}{2t_{2}^3 + 2\pi^2 \textit{a}^2 t_{2}}}{1- \frac{\pi^2 \textit{a}^3 }{2}\left[ \frac{\left(e^{\frac{2t_{1}}{\textit{a}}}-1\right)e^{-\frac{t_{1}}{\lambda_1}-\frac{t_{1}}{\textit{a}}}}{t_{1}^3 + \pi^2 \textit{a}^2 t_{1} }- \frac{\left(e^{\frac{2t_{2}}{\textit{a}}}-1\right)e^{-\frac{t_{2}}{\lambda_1}-\frac{t_{2}}{\textit{a}}}}{t_{2}^3 + \pi^2 \textit{a}^2 t_{2} }\right]} \stackrel{\sim}{\leq} \beta,
\label{2.11}
\end{aligned}
\end{equation}
where $t_{1}, t_{2}>0$, $t_{2}>t_{1}, \textrm{\textit{a}} > 0 \text{ and } \textrm{\textit{a}} > \lambda_j, j=0,1.$
Note that, the integrals in $E_0(Y_i)$ do not have a closed-form expression. Hence, we consider two different approaches of solving the problem $P_1'(t_1, t_2)$:
\begin{itemize}
    \item[] Case (i): First, we evaluate the integrals in $E_0(Y_i)$ numerically, by using the numerical integration method called composite Simpson's rule (see \citeN{richard2011}) and get an approximate value for $E_0(Y_i)$, say $E_0(Y_i)^*$. 
    \item[] Case (ii): We find an upper bound for $E_0(Y_i)$, and solve a new problem $Q_1(t_1, t_2)$, as follows:
   \end{itemize} 
Minimize \ $ETC_{UB}$ such that conditions~(\ref{2.9}) and~(\ref{2.10}) are satisfied, 
\begin{equation}
    \text{where,} \ ETC_{UB} = \frac{C \times E_0(Y_i)^{**}}{1-p_{c}}, 
    \label{2.13}
\end{equation}
and 
\begin{equation*}
  \left| E_0(Y_i)\right|  \leq \frac{\textrm{\textit{a}}}{2}  \operatorname{ln} \left(\frac{\textrm{\textit{a}}+\lambda_0}{\textrm{\textit{a}}-\lambda_0}\right) \left[ 1 + \left| \cos \left( \frac{\textrm{\textit{a}}\pi}{\lambda_0}\right) \right| + \left| \sin \left(\frac{\textrm{\textit{a}}\pi}{\lambda_0} \right) \right|  \right] = E_0(Y_i)^{**}.  
\end{equation*}
After the substitution of expressions for $p_{c}, p_{a}$ and $p_{r}$ from~(\ref{2.6}),~(\ref{2.7}) and~(\ref{2.8}), the fuzzy optimization problem, $Q_1(t_1, t_2)$, becomes problem $Q_1'(t_1, t_2)$:
\begin{equation*}
\begin{split}
Q_1'(t_1, t_2): \qquad \underset{\left(t_{1}, t_{2}\right)}{\operatorname{Min}}\quad  &\frac{C E_0(Y_i)^{**}}{1- \frac{\pi^2 \textit{a}^3 }{2}\left[ \frac{\left(e^{\frac{2t_{1}}{\textit{a}}}-1\right)e^{-\frac{t_{1}}{\lambda_0}-\frac{t_{1}}{\textit{a}}}}{t_{1}^3 + \pi^2 \textit{a}^2 t_{1} }- \frac{\left(e^{\frac{2t_{2}}{\textit{a}}}-1\right)e^{-\frac{t_{2}}{\lambda_0}-\frac{t_{2}}{\textit{a}}}}{t_{2}^3 + \pi^2 \textit{a}^2 t_{2} }\right]} \\
\text { such that} \quad&\frac{1 - \frac{\pi^2 \textit{a}^3 \left(e^{\frac{2t_{1}}{\textit{a}}}-1\right)e^{-\frac{t_{1}}{\lambda_0}-\frac{t_{1}}{\textit{a}}}}{2t_{1}^3 + 2\pi^2 \textit{a}^2 t_{1}}}{1- \frac{\pi^2 \textit{a}^3 }{2}\left[ \frac{\left(e^{\frac{2t_{1}}{\textit{a}}}-1\right)e^{-\frac{t_{1}}{\lambda_0}-\frac{t_{1}}{\textit{a}}}}{t_{1}^3 + \pi^2 \textit{a}^2 t_{1} }- \frac{\left(e^{\frac{2t_{2}}{\textit{a}}}-1\right)e^{-\frac{t_{2}}{\lambda_0}-\frac{t_{2}}{\textit{a}}}}{t_{2}^3 + \pi^2 \textit{a}^2 t_{2} }\right]} \stackrel{\sim}{\leq} \alpha,\\
&\frac{\frac{\pi^2 \textit{a}^3 \left(e^{\frac{2t_{2}}{\textit{a}}}-1\right)e^{-\frac{t_{2}}{\lambda_1}-\frac{t_{2}}{\textit{a}}}}{2t_{2}^3 + 2\pi^2 \textit{a}^2 t_{2}}}{1- \frac{\pi^2 \textit{a}^3 }{2}\left[ \frac{\left(e^{\frac{2t_{1}}{\textit{a}}}-1\right)e^{-\frac{t_{1}}{\lambda_1}-\frac{t_{1}}{\textit{a}}}}{t_{1}^3 + \pi^2 \textit{a}^2 t_{1} }- \frac{\left(e^{\frac{2t_{2}}{\textit{a}}}-1\right)e^{-\frac{t_{2}}{\lambda_1}-\frac{t_{2}}{\textit{a}}}}{t_{2}^3 + \pi^2 \textit{a}^2 t_{2} }\right]} \stackrel{\sim}{\leq} \beta,
\end{split}
\end{equation*}

where $t_{1}, t_{2}>0$, $t_{2}>t_{1}, \textrm{\textit{a}} > 0 \text{ and } \textrm{\textit{a}} > \lambda_j, j=0,1.$
\subsubsection{Fuzzy optimization algorithm}
\begin{figure}[!htb]
\centering
 \includegraphics[width=6cm,height=2.8cm]{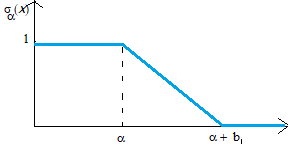}
 \caption{Membership function of $\alpha$.}
 \label{fig:Fig2}
\end{figure}

The optimization problem $P_1'(t_1, t_2)$ and $Q_1'(t_1, t_2)$ are  fuzzy nonlinear programming problems (NLPP) and they are converted to crisp nonlinear programming problems $P_1''(t_1, t_2)$ and $Q_1''(t_1, t_2)$, respectively by the following steps as in \citeN{jameel2012} with fuzzy parameters \ $\stackrel{\sim}{\alpha} \ = \left\{\left(x, \sigma_{\alpha}(x)\right) : x \in \mathbb{R}\right\}$ and $\stackrel{\sim}{\beta} \ = \left\{\left(x, \sigma_{\beta}(x)\right) : x \in \mathbb{R}\right\}$ having membership functions (see Figure~\ref{fig:Fig2}),
\begin{equation}
  \sigma_{\alpha}(x) = \begin{cases}
       1, & \text{if } x < \alpha, \\
       \frac{\alpha + b_1 - x}{b_1}, & \text{if } \alpha \leq x < \alpha + b_1, \\
       0, & \text{if } x \geq \alpha + b_1,
     \end{cases}  
     \label{2.14}
\end{equation}
\begin{equation}
  \sigma_{\beta}(x) = \begin{cases}
       1, & \text{if } x < \beta, \\
       \frac{\beta + b_2 - x}{b_2}, & \text{if } \beta \leq x < \beta + b_2, \\
       0, & \text{if } x \geq \beta + b_2,
     \end{cases}
     \label{2.15} 
\end{equation}
where $b_i \in [0, 1], i = 1, 2$.
\begin{enumerate}
    \item[] Step 1: First, solve the following NLPP, \\
    \begin{equation}
\begin{aligned}[b]
Z_1(t_1, t_2): \qquad &\underset{\left(t_{1}, t_{2}\right)}{\operatorname{Min}}\quad   \text{ETC*} \\
\text { such that} \qquad & g_1(t_1, t_2) \leq \alpha,\\
& h_1(t_1, t_2) \leq \beta,  \label{2.16}
\end{aligned}
\end{equation}
\begin{equation*}
    \begin{split}
    \text{where} \ \ \text{ETC*} &= \frac{C E_0(Y_i)^*}{1- \frac{\pi^2 \textit{a}^3 }{2}\left[ \frac{\left(e^{\frac{2t_{1}}{\textit{a}}}-1\right)e^{-\frac{t_{1}}{\lambda_0}-\frac{t_{1}}{\textit{a}}}}{t_{1}^3 + \pi^2 \textit{a}^2 t_{1} }- \frac{\left(e^{\frac{2t_{2}}{\textit{a}}}-1\right)e^{-\frac{t_{2}}{\lambda_0}-\frac{t_{2}}{\textit{a}}}}{t_{2}^3 + \pi^2 \textit{a}^2 t_{2} }\right]},\\
    g_1(t_1, t_2) &= \frac{1 - \frac{\pi^2 \textit{a}^3 \left(e^{\frac{2t_{1}}{\textit{a}}}-1\right)e^{-\frac{t_{1}}{\lambda_0}-\frac{t_{1}}{\textit{a}}}}{2t_{1}^3 + 2\pi^2 \textit{a}^2 t_{1}}}{1- \frac{\pi^2 \textit{a}^3 }{2}\left[ \frac{\left(e^{\frac{2t_{1}}{\textit{a}}}-1\right)e^{-\frac{t_{1}}{\lambda_0}-\frac{t_{1}}{\textit{a}}}}{t_{1}^3 + \pi^2 \textit{a}^2 t_{1} }- \frac{\left(e^{\frac{2t_{2}}{\textit{a}}}-1\right)e^{-\frac{t_{2}}{\lambda_0}-\frac{t_{2}}{\textit{a}}}}{t_{2}^3 + \pi^2 \textit{a}^2 t_{2} }\right]},
    \end{split}
\end{equation*}
\begin{equation*}
  h_1(t_1, t_2) = \frac{\frac{\pi^2 \textit{a}^3 \left(e^{\frac{2t_{2}}{\textit{a}}}-1\right)e^{-\frac{t_{2}}{\lambda_1}-\frac{t_{2}}{\textit{a}}}}{2t_{2}^3 + 2\pi^2 \textit{a}^2 t_{2}}}{1- \frac{\pi^2 \textit{a}^3 }{2}\left[ \frac{\left(e^{\frac{2t_{1}}{\textit{a}}}-1\right)e^{-\frac{t_{1}}{\lambda_1}-\frac{t_{1}}{\textit{a}}}}{t_{1}^3 + \pi^2 \textit{a}^2 t_{1} }- \frac{\left(e^{\frac{2t_{2}}{\textit{a}}}-1\right)e^{-\frac{t_{2}}{\lambda_1}-\frac{t_{2}}{\textit{a}}}}{t_{2}^3 + \pi^2 \textit{a}^2 t_{2} }\right]},    
\end{equation*}
 $t_{1}, t_{2}>0$, $t_{2}>t_{1}, \textrm{\textit{a}} > 0, \textrm{\textit{a}} > \lambda_j, j=0,1$ and $b_i \in [0, 1], i = 1, 2$.
 \item[] Step 2: Then, solve the NLPP,
  \begin{equation}
\begin{aligned}[b]
 Z_2(t_1, t_2): \qquad &\underset{\left(t_{1}, t_{2}\right)}{\operatorname{Min}}\quad   \text{ETC*} \\
\text { such that} \qquad & g_1(t_1, t_2) \leq \alpha + b_1 ,\\
& h_1(t_1, t_2) \leq \beta + b_2,  \label{2.17}\\
\end{aligned}
\end{equation}
where $t_{1}, t_{2}>0$, $t_{2}>t_{1}, \textrm{\textit{a}} > 0, \textrm{\textit{a}} > \lambda_j, j=0,1$ and $b_i \in [0, 1], i = 1, 2$.
\item[] Step 3: After solving the NLPPs given by~(\ref{2.16}) and~(\ref{2.17}), we get the optimal values for $Z_1(t_1, t_2)$ and $Z_2(t_1, t_2)$. Then, find
\begin{equation*}
\begin{split}
   Z_{U} &= \max \left\{Z_1(t_1, t_2), Z_2(t_1, t_2)\right\},\\
    Z_{L} &= \min \left\{Z_1(t_1, t_2), Z_2(t_1, t_2)\right\}.
\end{split}
\end{equation*}
\item[] Step 4: Finally, we get the corresponding crisp alternative of the fuzzy NLPP $P_1'(t_1, t_2)$ as
\begin{equation*}
\begin{split}
P_1''(t_1, t_2): \qquad &\underset{\left(t_{1}, t_{2}\right)} {\operatorname{Max}}\quad  \phi \\
\text { such that} \qquad &\phi - \left( \frac{\text{ETC*} - Z_L}{Z_U - Z_L}\right) \leq 0,\\
& \phi - \left( \frac{\alpha + b_1 - g_1(t_1, t_2)}{b_1}\right) \leq 0,\\
& \phi - \left( \frac{\beta + b_2 - h_1(t_1, t_2)}{b_2}\right) \leq 0,\\
\end{split}
\end{equation*}
where $t_{1}, t_{2}>0$, $t_{2}>t_{1}, \textrm{\textit{a}} > 0, \textrm{\textit{a}} > \lambda_j, j=0,1, b_i \in [0, 1], i = 1, 2$ and $ \phi \in [0, 1]$.
\item[] Step 5: Similarly, by interchanging ETC* in~(\ref{2.16}) and~(\ref{2.17}) with $ETC_{UB}$ given by~(\ref{2.13}) and by repeating the above Steps 1 to 4 with $ETC_{UB}$, we get the corresponding crisp NLPP, $Q_1''(t_1, t_2)$, of $Q_1'(t_1, t_2)$ given by,
\end{enumerate}
\vspace{-2mm}
\begin{equation*}
\begin{split}
Q_1''(t_1, t_2): \qquad &\underset{\left(t_{1}, t_{2}\right)}{\operatorname{Max}}\quad   \phi \\
\text { such that} \qquad &\phi - \left( \frac{\text{ETC}_{UB} - Z_L}{Z_U - Z_L}\right) \leq 0,\\
& \phi - \left( \frac{\alpha + b_1 - g_1(t_1, t_2)}{b_1}\right) \leq 0,\\
& \phi - \left( \frac{\beta + b_2 - h_1(t_1, t_2)}{b_2}\right) \leq 0,\\
\end{split}
\end{equation*}
where $t_{1}, t_{2}>0$, $t_{2}>t_{1}, \textrm{\textit{a}} > 0, \textrm{\textit{a}} > \lambda_j, j=0,1, b_i \in [0, 1], i = 1, 2$ and $ \phi \in [0, 1]$.\\
The optimal pair $(t_1, t_2)$ is tabulated in Table~\ref{tab:SSP} using fmincon solver in MATLAB in both of these methods for different values of fuzzy parameters. In Table~\ref{tab:SSPcrisp}, the results of \citeN{MK} is presented. Next, a comparative study is made using Table~\ref{tab:SSP} and \ref{tab:SSPcrisp}.  
\section{Repetitive Group Sampling Plan with Fuzzy Parameters}
In this section, we develop three different fuzzy acceptance sampling plans using repetitive group sampling plans (RGSP) based on minimum of $n$ samples, maximum of $n$ samples, and Type I censoring. The lifetime of units is assumed as exponentially distributed with pdf given by~(\ref{2.1}). The decision on rejection and acceptance of the lot is made using fuzzy inequalities given by~(\ref{2.9}) and~(\ref{2.10}) with fuzzy parameters AQL, RQL, $\alpha$, and $\beta$. In the following subsections, we describe these sampling plans, and the corresponding results are compared. 
\subsection{Acceptance sampling plan based on minimum of \textit{n} samples}
Here, the acceptance sampling plan is based on the time of the first failure of an item in a random sample of size $n$. Suppose independent and identically distributed random variables $X_i,\ i=1,2,3,\cdots,n$ denotes the lifetime of an $i^{th}$ item in a random sample of size $n$.\\ Let $Y_{min}=\min \left\{X_{1}, X_{2}, \ldots, X_{n}\right\}$ and it follows the exponential distribution with mean $\frac{1}{n\lambda}$.
The acceptance sampling plan procedure is as follows:
\begin{enumerate}
    \item Fix the values of AQL and RQL at $\alpha$ and $\beta$.
    \item Take a random sample of size $n$ and obtain $Y_{min}$.
    \item Reject the lot, if  $Y_{min}< t_1$ or accept the lot, if $Y_{min} \geq t_2$, else, move to Step 4.
\item If $t_1 \leq Y_{min} < t_2$, then Step 2 is repeated by taking a new sample from the lot.
\end{enumerate}
The aim is to compute the optimal value of $(n, t_1, t_2)$  which minimizes the cost of testing subject to the fuzzy hypothesis~(\ref{2.3}) with $\alpha$ and $\beta$ taken as fuzzy numbers. Under the hypothesis $H_j, j= 0, 1$, the weighted probability density function of $Y_{min}$ is defined by
\begin{align*}
\begin{split}
     f_{j}(y)=\biggr\{\left[\left(\textit{\textrm{a}}-\lambda_{j}\right) n^3 y^{3}+3 \textit{\textrm{a}} n^2 y^{2} \lambda_{j}+\pi^{2} \textit{\textrm{a}}^{2}\left(\textit{\textrm{a}}-\lambda_{j}\right) n y +\pi^{2} \textit{\textrm{a}}^{3} \lambda_{j}\right] e^{\frac{2 n y}{\textit{\textrm{a}}}}- (\lambda_j + \textrm{\textit{a}}) n^3 y^{3}\\
             -3 \textit{\textrm{a}} n^2 y^{2} \lambda_{j}-\pi^{2} \textit{\textrm{a}}^{2}(\textrm{\textit{a}}+\lambda_j) n y - \pi^{2} \textit{\textrm{a}}^{3} \lambda_{j}\biggr\} \frac{\pi^{2} \textit{\textrm{a}}^{2} e^{-\frac{ny}{\textit{\textrm{a}}}-\frac{ny}{\lambda}_{j}}}{2ny^2 \left(n^2y^2 + \pi^2 \textit{\textrm{a}} \right)^2 \lambda_j}.
     \end{split}
\end{align*}
The probability of acceptance, rejection of the lot, and continuation of the sampling process under the hypothesis $H_j, j=0,1$ are given respectively as:
\begin{equation}
  p_{\textit{\textit{a}}}=P\left(Y_{min} \geq t_{2}\right) = \frac{\pi^2 \textit{a}^3 \left(e^{\frac{2nt_{2}}{\textit{a}}}-1\right)e^{-\frac{nt_{2}}{\lambda_j}-\frac{nt_{2}}{\textit{a}}}}{2n^3t_{2}^3 + 2n  t_{2} \pi^2 \textit{a}^2},
  \label{3.1}
\end{equation}
\vspace{-2mm}
\begin{equation}
p_{r}=P\left(Y_{min}<t_{1}\right) = 1 - \frac{\pi^2 \textit{a}^3 \left(e^{\frac{2nt_{1}}{\textit{a}}}-1\right)e^{-\frac{nt_{1}}{\lambda_j}-\frac{nt_{1}}{\textit{a}}}}{2n^3t_{1}^3 + 2nt_1\pi^2 \textit{a}^2}, 
\label{3.2}
\end{equation}
\begin{equation}
p_{c}=P\left(t_{1} \leq Y_{min}<t_{2}\right) = \frac{\pi^2 \textit{a}^3 }{2n}\left[ \frac{\left(e^{\frac{2nt_{1}}{\textit{a}}}-1\right)e^{-\frac{nt_{1}}{\lambda_j}-\frac{nt_{1}}{\textit{a}}}}{n^2t_{1}^3 + \pi^2 \textit{a}^2 t_{1} }- \frac{\left(e^{\frac{2nt_{2}}{\textit{a}}}-1\right)e^{-\frac{nt_{2}}{\lambda_j}-\frac{nt_{2}}{\textit{a}}}}{n^2t_{2}^3 + \pi^2 \textit{a}^2 t_{2} }\right].
\label{3.3}
\end{equation}
Note that, the total cost of testing is the product of the total testing time and the testing cost of an item for unit time. But the total time for concluding the testing process is the random variable since $Y_{min}$ and sample number are random variables. Hence, we compute the expected testing cost similar to the previous section by assuming $C$ as the testing cost of an item for unit time. Let $E_j(Y_{min})$ denotes the expectation of $Y_{min}$ under $H_j, j=0, 1.$ Then expected testing cost (ETC) is given by,
\begin{equation}
\mathrm{ETC}=\frac{C \times E_j(Y_{min})}{1-p_{c}}.
\label{3.4}
\end{equation}
We observe that minimizing the random total cost subject to constraints~(\ref{2.9}) and~(\ref{2.10}) will lead to an intractable optimization problem. Hence, in this case, we minimize ETC subjecting to the constraints~(\ref{2.9}) and~(\ref{2.10}) and obtain optimal values of $(n, t_1, t_2)$. The next subsection will describe the formulation of the fuzzy optimization problem and its solution procedure.
\subsubsection{Fuzzy optimization problem and solution procedure}
The expectation of $Y_{min}$ under null hypothesis $H_0$ is taken for calculating expected cost since it is a function of unknown parameter $\lambda$. Thus, the expectation of $Y_{min}$ under $H_0$ is given by,
\begin{equation*} 
   E_0(Y_{min}) = \frac{\textrm{\textit{a}}}{2n} \left[\operatorname{ln} \left[\frac{\textrm{\textit{a}}+\lambda_0}{\textrm{\textit{a}}-\lambda_0}\right]+
 \cos \left( \frac{\textrm{\textit{a}}\pi}{\lambda_0}\right) \displaystyle\int\limits_{ \frac{\textrm{\textit{a}}\pi}{\lambda_0}-\pi}^{ \frac{\textrm{\textit{a}}\pi}{\lambda_0}+\pi} \frac{\cos u}{u} d u+\sin \left(\frac{\textrm{\textit{a}}\pi}{\lambda_0} \right) \displaystyle\int\limits_{\frac{\textrm{\textit{a}}\pi}{\lambda_0} -\pi}^{\frac{\textrm{\textit{a}}\pi}{\lambda_0}+\pi} \frac{\sin u}{u} d u\right].    
\end{equation*}
It can be seen that the integrals in the above equation do not have an elementary anti-derivative without which one cannot proceed to solve the optimization problem. Hence, we resort to numerical integration to compute approximate value for $E_0(Y_{min})$. Composite Simpson's rule is applied to obtain $E_0(Y_{min})$ numerically (say, $E_0(Y_{min})^*$). Thus, we form the following nonlinear fuzzy optimization problem $P_2(n, t_1, t_2)$ given by 
\begin{equation} 
\begin{aligned}[b]
P_2(n, t_1, t_2):\qquad &\underset{\left(n,t_{1}, t_{2}\right)}{\operatorname{Min}}\left(\frac{C E_0(Y_{min})^*}{1-p_{c}}\right) \quad \text {such that}\\
 &P\left( \text{Reject the lot}  \mid \lambda \approx \frac{1}{\lambda_0}\right) \stackrel{\sim}{\leq} \alpha,  \label{3.5}
\end{aligned}
\end{equation}
\begin{equation} 
\qquad \qquad \qquad \quad P\left(\text {Accept the lot} \mid \lambda \approx \frac{1}{\lambda_{1}}\right) \stackrel{\sim}{\leq} \beta, 
\label{3.6}
\end{equation}
where $t_{1}, t_{2}>0, t_{2}>t_{1}$ and $\lambda_0 \geq t_2$ (by the definition of ASP). That is,
\begin{equation*}
\begin{split}
&\underset{\left(n,t_{1}, t_{2}\right)}{\operatorname{Min}}\left(\frac{C E_0(Y_{min})^*}{1-p_{c}}\right) \\
\text { such that} \qquad &\left(\frac{p_{r}}{1-p_{c}} \mid \lambda \approx \frac{1}{\lambda_{0}}\right) \stackrel{\sim}{\leq} \alpha, \\
&\left(\frac{p_{a}}{1-p_{c}} \mid \lambda \approx \frac{1}{\lambda_{1}}\right) \stackrel{\sim}{\leq} \beta,
\end{split}
\end{equation*}
where $t_{1}, t_{2}>0$ and $t_{2}>t_{1}$. Substituting for $p_{a}$, $p_{r}$ and $p_{c}$ from Eqs~(\ref{3.1}),~(\ref{3.2}) and~(\ref{3.3}), the problem $P_2(n, t_1, t_2)$ becomes $P_2'(n, t_1, t_2)$,
\begin{equation}
\begin{aligned}[b]
P_2'(n, t_1, t_2): \qquad &\underset{\left(n,t_{1}, t_{2}\right)}{\operatorname{Min}} \text{ETC}^* \\
\text{such that} \qquad & g_2(n, t_1, t_2) \stackrel{\sim}{\leq} \alpha, \\
&h_2(n, t_1, t_2) \stackrel{\sim}{\leq} \beta, \label{3.7}
\end{aligned}
\end{equation}
\begin{equation*}
  \text{where} \ \ \text{ETC}^* = \frac{C E_0(Y_{min})^*}{1- \frac{\pi^2 \textit{a}^3 }{2n}\left[ \frac{\left(e^{\frac{2nt_{1}}{\textit{a}}}-1\right)e^{-\frac{nt_{1}}{\lambda_0}-\frac{nt_{1}}{\textit{a}}}}{n^2t_{1}^3 + \pi^2 \textit{a}^2 t_{1} }- \frac{\left(e^{\frac{2nt_{2}}{\textit{a}}}-1\right)e^{-\frac{nt_{2}}{\lambda_0}-\frac{nt_{2}}{\textit{a}}}}{n^2t_{2}^3 + \pi^2 \textit{a}^2 t_{2} }\right]},  
\end{equation*}
\begin{equation*}
    \begin{split}
        g_2(n, t_1, t_2)&= \frac{1 - \frac{\pi^2 \textit{a}^3 \left(e^{\frac{2nt_{1}}{\textit{a}}}-1\right)e^{-\frac{nt_{1}}{\lambda_0}-\frac{nt_{1}}{\textit{a}}}}{2n^3t_{1}^3 + 2nt_{1}\pi^2 \textit{a}^2}}{1- \frac{\pi^2 \textit{a}^3 }{2n}\left[ \frac{\left(e^{\frac{2nt_{1}}{\textit{a}}}-1\right)e^{-\frac{nt_{1}}{\lambda_0}-\frac{nt_{1}}{\textit{a}}}}{n^2t_{1}^3 + \pi^2 \textit{a}^2 t_{1} }- \frac{\left(e^{\frac{2nt_{2}}{\textit{a}}}-1\right)e^{-\frac{nt_{2}}{\lambda_0}-\frac{nt_{2}}{\textit{a}}}}{n^2t_{2}^3 + \pi^2 \textit{a}^2 t_{2} }\right]},\\
        h_2(n, t_1, t_2)&=\frac{\frac{\pi^2 \textit{a}^3 \left(e^{\frac{2nt_{2}}{\textit{a}}}-1\right)e^{-\frac{nt_{2}}{\lambda_1}-\frac{nt_{2}}{\textit{a}}}}{2n^3t_{2}^3 + 2nt_{2}\pi^2 \textit{a}^2}}{1- \frac{\pi^2 \textit{a}^3 }{2n}\left[ \frac{\left(e^{\frac{2nt_{1}}{\textit{a}}}-1\right)e^{-\frac{nt_{1}}{\lambda_1}-\frac{nt_{1}}{\textit{a}}}}{n^2t_{1}^3 + \pi^2 \textit{a}^2 t_{1} }- \frac{\left(e^{\frac{2nt_{2}}{\textit{a}}}-1\right)e^{-\frac{nt_{2}}{\lambda_1}-\frac{nt_{2}}{\textit{a}}}}{n^2t_{2}^3 + \pi^2 \textit{a}^2 t_{2} }\right]}, 
    \end{split}
\end{equation*}
$t_{1}, t_{2}>0$, $t_{2}>t_{1}, \textrm{\textit{a}} > 0 \text{ and } \textrm{\textit{a}} > \lambda_j, j=0,1.$\\
The membership functions given in~(\ref{2.14}) and~(\ref{2.15}) are used for the parameters $\alpha$ and $\beta$ in $P_2'(n, t_1, t_2)$.
Next, the crisp nonlinear optimization problem, $P_2''(n, t_1, t_2)$, corresponding to $P_2'(n, t_1, t_2)$ is obtained by applying the steps of the fuzzy optimization algorithm discussed in Subsection 2.1.1 to the problem $P_2'(n, t_1, t_2)$ and $P_2''(n, t_1, t_2)$ is given by,
\begin{equation*}
\begin{split}
P_2''(n, t_1, t_2): \qquad &\underset{\left(n, t_{1}, t_{2}\right)} {\operatorname{Max}}\quad  \phi \\
\text { such that} \qquad &\phi - \left( \frac{\text{ETC*} - Z_L}{Z_U - Z_L}\right) \leq 0\\
& \phi - \left( \frac{\alpha + b_1 - g_2(n, t_1, t_2)}{b_1}\right) \leq 0,\\
& \phi - \left( \frac{\beta + b_2 - h_2(n, t_1, t_2)}{b_2}\right) \leq 0,\\
\end{split}
\end{equation*}
$\text{where} \quad Z_{U} = \max \left\{Z_1(n, t_1, t_2), Z_2(n, t_1, t_2)\right\},\
    Z_{L} = \min \left\{Z_1(n, t_1, t_2), Z_2(n, t_1, t_2)\right\},$
$t_{1}, t_{2}>0$, $t_{2}>t_{1}, \textrm{\textit{a}} > 0, \ \textrm{\textit{a}} > \lambda_j, j=0,1, \ b_i \in [0, 1], i = 1, 2$ and $ \phi \in [0, 1]$.

The optimal solution of $P_2''(n, t_1, t_2)$ is obtained by solving  the nonlinear optimization problem using fmincon solver in MATLAB and results are tabulated in Table~\ref{tab:min} for various values of $\textit{a}, b_1, b_2, \lambda_1, \lambda _2, \alpha \text{ and } \beta$ and compared with crisp case given in Table~\ref{tab:mincrisp}.

Observe that the problem $P_2'(n, t_1, t_2)$ consists of a objective function ETC, which involves $E_0(Y_{min})$. Note that this $E_0(Y_{min})$ is approximated by replacing the integrals in it with the numerical equivalent obtained by the composite Simpson's rule. Next, we consider another problem $Q_2(n, t_1, t_2)$ of minimizing the upper bound for ETC subject to constraints~(\ref{3.5}) and~(\ref{3.6}). This will again form a fuzzy optimization problem where the aim is to find optimal values for $n$, $t_1$, and $t_2$. It is clear that the actual total cost will be lower than the upper bound for ETC ($ETC_{UB}$), where $ETC_{UB}$ is given by,
\begin{equation}
     ETC_{UB} = \frac{C \times E_0(Y_{min})^{**}}{1-p_{c}}  
     \label{3.9}
\end{equation}
\begin{equation*}
  \text{and} \qquad \left| E_0(Y_{min})\right|  \leq \frac{\textrm{\textit{a}}}{2n}  \operatorname{ln} \left(\frac{\textrm{\textit{a}}+\lambda_0}{\textrm{\textit{a}}-\lambda_0}\right) \left[ 1 + \left| \cos \left( \frac{\textrm{\textit{a}}\pi}{\lambda_0}\right) \right| + \left| \sin \left(\frac{\textrm{\textit{a}}\pi}{\lambda_0} \right) \right|  \right] = E_0(Y_{min})^{**}.  
\end{equation*}
Thus, we consider the following fuzzy optimization problems:
\begin{equation*}
\begin{split}
Q_2(n, t_1, t_2): \qquad &\underset{\left(n,t_{1}, t_{2}\right)}{\operatorname{Min}}\left(\frac{C E_0(Y_{min})^{**}}{1-p_{c}}\right)\\
 \text { such that} \qquad &\left(\frac{p_{r}}{1-p_{c}} \mid \lambda \approx \frac{1}{\lambda_{0}}\right) \stackrel{\sim}{\leq} \alpha, \\
&\left(\frac{p_{a}}{1-p_{c}} \mid \lambda \approx \frac{1}{\lambda_{1}}\right) \stackrel{\sim}{\leq} \beta,
\end{split}
\end{equation*}
where $t_{1}, t_{2}>0$ and $t_{2}>t_{1}$. Substituting for $p_{a}$, $p_{r}$ and $p_{c}$ from Eqs~(\ref{3.1}),~(\ref{3.2}) and~(\ref{3.3}), the problem $Q_2(n, t_1, t_2)$ becomes $Q_2'(n, t_1, t_2)$,
\begin{equation*}
\begin{split}
Q_2'(n, t_1, t_2): \qquad &\underset{\left(n,t_{1}, t_{2}\right)}{\operatorname{Min}}\frac{C E_0(Y_{min})^{**}}{1- \frac{\pi^2 \textit{a}^3 }{2n}\left[ \frac{\left(e^{\frac{2nt_{1}}{\textit{a}}}-1\right)e^{-\frac{nt_{1}}{\lambda_0}-\frac{nt_{1}}{\textit{a}}}}{n^2t_{1}^3 + \pi^2 \textit{a}^2 t_{1} }- \frac{\left(e^{\frac{2nt_{2}}{\textit{a}}}-1\right)e^{-\frac{nt_{2}}{\lambda_0}-\frac{nt_{2}}{\textit{a}}}}{n^2t_{2}^3 + \pi^2 \textit{a}^2 t_{2} }\right]} \\
\text { such that} \qquad & g_2(n, t_1, t_2) \stackrel{\sim}{\leq} \alpha,\\
& h_2(n, t_1, t_2) \stackrel{\sim}{\leq} \beta,
\end{split}
\end{equation*}
where $t_{1}, t_{2}>0$, $t_{2}>t_{1}, \textrm{\textit{a}} > 0 \text{ and } \textrm{\textit{a}} > \lambda_j, j=0,1.$

The fuzzy nonlinear optimization problem, $Q_2'(n, t_1, t_2)$, is converted to the corresponding crisp nonlinear optimization problem, by following the fuzzy optimization algorithm in Subsection 2.1.1 for  $Q_2'(n, t_1, t_2)$. The crisp nonlinear problem, $Q_2''(n, t_1, t_2)$, hence obtained is given by,
\begin{equation*}
\begin{split}
Q_2''(n, t_1, t_2): \qquad &\underset{\left(n, t_{1}, t_{2}\right)} {\operatorname{Max}}\quad  \phi \\
\text { such that} \qquad &\phi - \left( \frac{\text{ETC}_{UB} - Z_L}{Z_U - Z_L}\right) \leq 0,\\
& \phi - \left( \frac{\alpha + b_1 - g_2(n, t_1, t_2)}{b_1}\right) \leq 0,\\
& \phi - \left( \frac{\beta + b_2 - h_2(n, t_1, t_2)}{b_2}\right) \leq 0,\\
\end{split}
\end{equation*}
\begin{equation*}
\begin{split}
\text{where}\ \ \text{ETC}_{UB} &= \frac{C E_0(Y_{min})^{**}}{1- \frac{\pi^2 \textit{a}^3 }{2n}\left[ \frac{\left(e^{\frac{2nt_{1}}{\textit{a}}}-1\right)e^{-\frac{nt_{1}}{\lambda_0}-\frac{nt_{1}}{\textit{a}}}}{n^2t_{1}^3 + \pi^2 \textit{a}^2 t_{1} }- \frac{\left(e^{\frac{2nt_{2}}{\textit{a}}}-1\right)e^{-\frac{nt_{2}}{\lambda_0}-\frac{nt_{2}}{\textit{a}}}}{n^2t_{2}^3 + \pi^2 \textit{a}^2 t_{2} }\right]},
\end{split}
\end{equation*}
$Z_{U} = \max \left\{Z_1(n, t_1, t_2), Z_2(n, t_1, t_2)\right\},\
    Z_{L} = \min \left\{Z_1(n, t_1, t_2), Z_2(n, t_1, t_2)\right\},$
$t_{1}, t_{2}>0$, $t_{2}>t_{1}, \textrm{\textit{a}}> 0, \ \textrm{\textit{a}} > \lambda_j, j=0,1, b_i \in [0, 1], i = 1, 2$ and $ \phi \in [0, 1]$.

The crisp nonlinear problem $Q_2''(n, t_1, t_2)$, is solved using fmincon solver in MATLAB and optimal values of $n, t_1$ and $t_2$ are obtained. Results are tabulated in Table~\ref{tab:min} for different values of $\textit{a}, b_1, b_2, \lambda_1, \lambda _2, \alpha \text{ and } \beta$ and compared with crisp case given in Table~\ref{tab:mincrisp}.
\subsection{Acceptance sampling plan based on maximum of n samples}
This section deals with the maximum lifetime among the $n$ samples taken at a time. Let $X_i, i = 1, 2, 3, \cdots ,n$ be the independent and identically distributed exponential random variables which denotes the lifetime of $i^{th}$ item with p.d.f given by~(\ref{2.1}) and $Y_{max} = \max (X_1,X_2,\cdots,X_n)$.  $P(Y_{max}>t) = 1 - P(Y_{max} \leq t) = 1 - \left(1-e^{-t\lambda}\right)^{n} $.
The acceptance sampling plan procedure (see \citeN{MK}) is as follows:
\begin{enumerate}
    \item Fix the values of AQL and RQL at specified producer's risk $\alpha$ and consumer's risk $\beta$.
\item Take a random sample of size $n$ and find the value of  $Y_{max}$.
\item Reject the lot, if  $Y_{max} < t_1$ or accept the lot, if $Y_{max} \geq t_2$, else, move to Step 4.
\item If $t_1 \leq Y_{max} < t_2$, then Step 2 is repeated by taking a new sample from the lot.
\end{enumerate}
As per our plan, the goal is to find the optimal values of $(n, t_1,t_2)$ which minimizes the total testing cost subject to the fuzzy hypothesis~(\ref{2.3}) with all the parameters, namely, AQL, RQL, $\alpha$ and $\beta$ are fuzzy numbers. We get the required probabilities, probability of continuation of the sampling process, probability of acceptance and rejection of lot under the hypothesis, $H_j$ for the given acceptance sampling plan respectively as,
\begin{equation}
 p_{c}=P\left(t_{1} \leq Y_{max} < t_{2}\right) = \left(1 - \frac{\pi^2 \textit{a}^3 \left(e^{\frac{2t_{2}}{\textit{a}}}-1\right)e^{-\frac{t_{2}}{\lambda_j}-\frac{t_{2}}{\textit{a}}}}{2t_{2}^3 + 2\pi^2 \textit{a}^2 t_{2}}\right)^n-\left(1 - \frac{\pi^2 \textit{a}^3 \left(e^{\frac{2t_{1}}{\textit{a}}}-1\right)e^{-\frac{t_{1}}{\lambda_j}-\frac{t_{1}}{\textit{a}}}}{2t_{1}^3 + 2\pi^2 \textit{a}^2 t_{1}}\right)^n, \label{3.10}
 \end{equation}
 \vspace{-2mm}
 \begin{equation}
p_{\textit{\textit{a}}}=P\left(Y_{max} \geq t_{2}\right) = 1- \left(1 - \frac{\pi^2 \textit{a}^3 \left(e^{\frac{2t_{2}}{\textit{a}}}-1\right)e^{-\frac{t_{2}}{\lambda_j}-\frac{t_{2}}{\textit{a}}}}{2t_{2}^3 + 2\pi^2 \textit{a}^2 t_{2}}\right)^n, 
\label{3.11}
 \end{equation}
 \begin{equation}
p_{r}=P\left(Y_{max}<t_{1}\right) = \left(1 - \frac{\pi^2 \textit{a}^3 \left(e^{\frac{2t_{1}}{\textit{a}}}-1\right)e^{-\frac{t_{1}}{\lambda_j}-\frac{t_{1}}{\textit{a}}}}{2t_{1}^3 + 2\pi^2 \textit{a}^2 t_{1}}\right)^n. 
\label{3.12}
 \end{equation}
Here, the overall testing cost is the product of the cost of testing a sample for unit time (C) and the total testing time. Note that, $Y_{max}$ and the number of samples are random quantities and hence the total testing time is a random quantity. Therefore, the expected testing cost (ETC) is calculated as the total testing cost is a random quantity. Let $E_j(Y_{max})$ denotes the expectation of $Y_{max}$ under the hypothesis $H_j, j = 0,1.$. Then the expected testing cost (ETC) is given by, 
\begin{equation}
   \mathrm{ETC}=\frac{C \times E_j(Y_{max})}{1-p_{c}}. 
   \label{3.13}
\end{equation}
 We note that the minimization of random costs subject to~(\ref{2.9}) and~(\ref{2.10}) leads to an intractable optimization problem. Thus, we minimize ETC such that conditions~(\ref{2.9}) and~(\ref{2.10}) are satisfied to obtain the optimal values of $(n, t_1, t_2)$. The fuzzy optimization problem formation and solution procedure are explained in the next subsection.
\subsubsection{Fuzzy optimization problem 
and solution procedure}
The expectation of $Y_{max}$ under the null hypothesis $H_0$ ($E_0(Y_{max})$) is obtained since ETC is a function of the unknown quantity $\lambda$ and is given by, 
\begin{equation*}
   E_0(Y_{max}) = \frac{\textrm{\textit{a}}}{2} \sum\limits_{i=1}^n \frac{1}{i} \left[\operatorname{ln} \left[\frac{\textrm{\textit{a}}+\lambda_0}{\textrm{\textit{a}}-\lambda_0}\right]+
 \cos \left( \frac{\textrm{\textit{a}}\pi}{\lambda_0}\right) \displaystyle\int\limits_{ \frac{\textrm{\textit{a}}\pi}{\lambda_0}-\pi}^{ \frac{\textrm{\textit{a}}\pi}{\lambda_0}+\pi} \frac{\cos u}{u} d u+\sin \left(\frac{\textrm{\textit{a}}\pi}{\lambda_0} \right) \displaystyle\int\limits_{\frac{\textrm{\textit{a}}\pi}{\lambda_0} -\pi}^{\frac{\textrm{\textit{a}}\pi}{\lambda_0}+\pi} \frac{\sin u}{u} d u\right].    
\end{equation*}
Hence, we consider the fuzzy optimization problem of minimizing ETC under $H_0$. Note that, the integrals in the equation of $E_0(Y_{max})$ do not have a closed-form expression. In order to solve the optimization problem, the integrals in $E_0(Y_{max})$ are computed numerically using composite Simpson's rule and obtain $E_O(Y_{max})^*$. Thus, we formulate the following nonlinear fuzzy optimization problem $P_3(n, t_1, t_2)$ given by,
\begin{equation} 
\begin{aligned}[b]
P_3(n, t_1, t_2):\qquad &\underset{\left(n,t_{1}, t_{2}\right)}{\operatorname{Min}}\left(\frac{C E_0(Y_{max})^*}{1-p_{c}}\right) \qquad \text {such that}\\
 &P\left( \text{Reject the lot}  \mid \lambda \approx \frac{1}{\lambda_0}\right) \stackrel{\sim}{\leq} \alpha, 
\end{aligned}
\label{3.14}
\end{equation}
\vspace{-10mm}
\begin{equation}
\qquad \qquad \qquad  P\left(\text {Accept the lot} \mid \lambda \approx \frac{1}{\lambda_{1}}\right) \stackrel{\sim}{\leq} \beta, 
\label{3.15}
\end{equation}
where $t_{1}, t_{2}>0$, $t_{2}>t_{1}$, and $t_2 \leq \lambda_0$. That is,
\begin{equation*}
\begin{split}
&\underset{\left(n,t_{1}, t_{2}\right)}{\operatorname{Min}}\left(\frac{C E_0(Y_{max})^*}{1-p_{c}}\right)\\
\text {such that} \qquad &\left(\frac{p_{r}}{1-p_{c}} \mid \lambda \approx \frac{1}{\lambda_{0}}\right) \stackrel{\sim}{\leq} \alpha, \\
&\left(\frac{p_{a}}{1-p_{c}} \mid \lambda \approx \frac{1}{\lambda_{1}}\right) \stackrel{\sim}{\leq} \beta,\qquad \qquad \qquad \qquad 
\end{split}
\end{equation*}
where $t_{1}, t_{2}>0$, $t_{2}>t_{1}$, and $t_2 \leq \lambda_0$. Substituting for $p_{c}$, $p_{a}$ and $p_{r}$ from~(\ref{3.10}),~(\ref{3.11}) and~(\ref{3.12}), the problem $P_3(n, t_1, t_2)$ becomes $P_3'(n, t_1, t_2)$,
\begin{equation}
\begin{aligned}[b]
P_3'(n, t_1, t_2): \qquad
&\underset{\left(n,t_{1}, t_{2}\right)}{\operatorname{Min}} \quad \text{ETC}^*\\
\text {such that} \qquad & g_3(n, t_1, t_2) \stackrel{\sim}{\leq} \alpha,\\
& h_3(n, t_1, t_2) \stackrel{\sim}{\leq} \beta,
\end{aligned}
\label{3.16}
\end{equation}
where
\begin{equation*}
    \begin{split}
     \text{ETC}^* &= \frac{C E_0(Y_{max})^*}{\left(1 - \frac{\pi^2 \textit{a}^3 \left(e^{\frac{2t_{2}}{\textit{a}}}-1\right)e^{-\frac{t_{2}}{\lambda_0}-\frac{t_{2}}{\textit{a}}}}{2t_{2}^3 + 2\pi^2 \textit{a}^2 t_{2}}\right)^n-\left(1 - \frac{\pi^2 \textit{a}^3 \left(e^{\frac{2t_{1}}{\textit{a}}}-1\right)e^{-\frac{t_{1}}{\lambda_0}-\frac{t_{1}}{\textit{a}}}}{2t_{1}^3 + 2\pi^2 \textit{a}^2 t_{1}}\right)^n},  \\
     g_3(n, t_1, t_2)&=\frac{\left(1 - \frac{\pi^2 \textit{a}^3 \left(e^{\frac{2t_{1}}{\textit{a}}}-1\right)e^{-\frac{t_{1}}{\lambda_0}-\frac{t_{1}}{\textit{a}}}}{2t_{1}^3 + 2\pi^2 \textit{a}^2 t_{1}}\right)^n}{\left(1 - \frac{\pi^2 \textit{a}^3 \left(e^{\frac{2t_{2}}{\textit{a}}}-1\right)e^{-\frac{t_{2}}{\lambda_0}-\frac{t_{2}}{\textit{a}}}}{2t_{2}^3 + 2\pi^2 \textit{a}^2 t_{2}}\right)^n-\left(1 - \frac{\pi^2 \textit{a}^3 \left(e^{\frac{2t_{1}}{\textit{a}}}-1\right)e^{-\frac{t_{1}}{\lambda_0}-\frac{t_{1}}{\textit{a}}}}{2t_{1}^3 + 2\pi^2 \textit{a}^2 t_{1}}\right)^n},
     \end{split}
\end{equation*}
\begin{equation*}
    h_3(n, t_1, t_2)=\frac{1- \left(1 - \frac{\pi^2 \textit{a}^3 \left(e^{\frac{2t_{2}}{\textit{a}}}-1\right)e^{-\frac{t_{2}}{\lambda_1}-\frac{t_{2}}{\textit{a}}}}{2t_{2}^3 + 2\pi^2 \textit{a}^2 t_{2}}\right)^n }{\left(1 - \frac{\pi^2 \textit{a}^3 \left(e^{\frac{2t_{2}}{\textit{a}}}-1\right)e^{-\frac{t_{2}}{\lambda_1}-\frac{t_{2}}{\textit{a}}}}{2t_{2}^3 + 2\pi^2 \textit{a}^2 t_{2}}\right)^n-\left(1 - \frac{\pi^2 \textit{a}^3 \left(e^{\frac{2t_{1}}{\textit{a}}}-1\right)e^{-\frac{t_{1}}{\lambda_1}-\frac{t_{1}}{\textit{a}}}}{2t_{1}^3 + 2\pi^2 \textit{a}^2 t_{1}}\right)^n},
\end{equation*}

$t_{1}, t_{2}>0$, $t_{2}>t_{1}, \textrm{\textit{a}} > 0, t_2 \leq \lambda_0, \text{ and } \textrm{\textit{a}} > \lambda_j, j=0,1.$

The membership functions of the fuzzy parameters $\alpha$ and $\beta$ in $P_3'(n, t_1, t_2)$ are taken as in~(\ref{2.14}) and~(\ref{2.15}). The steps of the fuzzy optimization algorithm described in the Subsection 2.1.1 are applied to the fuzzy nonlinear optimization problem $P_3'(n, t_1, t_2)$ and we get the corresponding crisp nonlinear optimization problem $P_3''(n, t_1, t_2)$, given by,
\begin{equation*}
\begin{split}
P_3''(n, t_1, t_2): \qquad &\underset{\left(n, t_{1}, t_{2}\right)} {\operatorname{Max}}\quad  \phi \\
\text { such that} \qquad &\phi - \left( \frac{\text{ETC*} - Z_L}{Z_U - Z_L}\right) \leq 0,\\
& \phi - \left( \frac{\alpha + b_1 - g_3(n, t_1, t_2)}{b_1}\right) \leq 0,\\
& \phi - \left( \frac{\beta + b_2 - h_3(n, t_1, t_2)}{b_2}\right) \leq 0,\\
\end{split}
\end{equation*}
 $ \text{where} \ Z_{U} = \max \left\{Z_1(n, t_1, t_2), Z_2(n, t_1, t_2)\right\},\
    Z_{L} = \min \left\{Z_1(n, t_1, t_2), Z_2(n, t_1, t_2)\right\},$
$t_{1}, t_{2}>0$, $t_{2}>t_{1},\ \textrm{\textit{a}} > 0,\ t_2 \leq \lambda_0,\ \phi \in [0, 1],\ \textrm{\textit{a}} > \lambda_j, j=0,1$, and $b_i \in [0, 1], i = 1, 2$.

The above crisp nonlinear optimization problem is solved to get the required optimal value of $(n, t_1, t_2)$ using the fmincon solver in MATLAB, and examples for different values of parameters in case of both crisp and fuzzy values are tabulated in Table~\ref{tab:max} and Table~\ref{tab:maxcrisp}, respectively to verify the computation.

 Next, we consider another optimization problem of minimizing an upper bound for ETC subject to the constraints~(\ref{3.14}) and~(\ref{3.15}) because of the presence of integrals with no closed-form expression in the objective function ETC. Hence, we formulate an optimization problem $Q_3(n, t_1, t_2)$ of minimizing the upper bound for ETC such that the fuzzy inequalities in~(\ref{3.14}) and~(\ref{3.15}) are satisfied and our aim is to get the optimal values for $n$, $t_1$ and $t_2$. Note that, the overall cost of testing will be less than the cost obtained through ETC upper bound ($ETC_{UB}$), where $ETC_{UB}$ is derived as,
 \begin{equation}
     ETC_{UB} = \frac{C \times E_0(Y_{max})^{**}}{1-p_{c}} 
     \label{3.18}
\end{equation}
\begin{equation*}
 \text{and} \qquad \left| E_0(Y_{max})\right|  \leq \frac{\textrm{\textit{a}}}{2}\sum\limits_{i=1}^n \frac{1}{i}   \operatorname{ln} \left(\frac{\textrm{\textit{a}}+\lambda_0}{\textrm{\textit{a}}-\lambda_0}\right) \left[ 1 + \left| \cos \left( \frac{\textrm{\textit{a}}\pi}{\lambda_0}\right) \right| + \left| \sin \left(\frac{\textrm{\textit{a}}\pi}{\lambda_0} \right) \right|  \right] = E_0(Y_{max})^{**}.  
\end{equation*}
Then the corresponding fuzzy nonlinear optimization problem is given by,
\begin{equation*}
\begin{split}
Q_3(n, t_1, t_2): \qquad &\underset{\left(n,t_{1}, t_{2}\right)}{\operatorname{Min}}\left(\frac{C E_0(Y_{max})^{**}}{1-p_{c}}\right) \\
\text {such that} \qquad &\left(\frac{p_{r}}{1-p_{c}} \mid \lambda \approx \frac{1}{\lambda_{0}}\right) \stackrel{\sim}{\leq} \alpha, \\
&\left(\frac{p_{a}}{1-p_{c}} \mid \lambda \approx \frac{1}{\lambda_{1}}\right) \stackrel{\sim}{\leq} \beta,
\end{split}
\end{equation*}

where $t_{1}, t_{2}>0$, $t_{2}>t_{1}$, and $t_2 \leq \lambda_0$. Substituting for $p_{c}$, $p_{a}$ and $p_{r}$ from~(\ref{3.10}),~(\ref{3.11}) and~(\ref{3.12}), the problem $Q_3(n, t_1, t_2)$ becomes $Q_3'(n, t_1, t_2)$,
\begin{equation}
\begin{aligned}[b]
Q_3'(n, t_1, t_2): \qquad
&\underset{\left(n,t_{1}, t_{2}\right)}{\operatorname{Min}} \quad \text{ETC}_{UB}\\
\text {such that} \qquad & g_3(n, t_1, t_2) \stackrel{\sim}{\leq} \alpha,\\
& h_3(n, t_1, t_2) \stackrel{\sim}{\leq} \beta,
\label{3.19}
\end{aligned}
\end{equation}
\begin{equation*}
     \textrm{where}\ \ \text{ETC}_{UB} =  \frac{C E_0(Y_{max})^{**}}{\left(1 - \frac{\pi^2 \textit{a}^3 \left(e^{\frac{2t_{2}}{\textit{a}}}-1\right)e^{-\frac{t_{2}}{\lambda_0}-\frac{t_{2}}{\textit{a}}}}{2t_{2}^3 + 2\pi^2 \textit{a}^2 t_{2}}\right)^n-\left(1 - \frac{\pi^2 \textit{a}^3 \left(e^{\frac{2t_{1}}{\textit{a}}}-1\right)e^{-\frac{t_{1}}{\lambda_0}-\frac{t_{1}}{\textit{a}}}}{2t_{1}^3 + 2\pi^2 \textit{a}^2 t_{1}}\right)^n},\qquad \qquad \qquad \qquad \qquad
      \end{equation*}
 $t_{1}, t_{2}>0$, $t_{2}>t_{1}, \textrm{\textit{a}} > 0, t_2 \leq \lambda_0 \text{ and } \textrm{\textit{a}} > \lambda_j, j=0,1.$
 
 By following the steps of the fuzzy optimization algorithm in Subsection 2.1.1 for $Q_3'(n, t_1, t_2)$, the fuzzy nonlinear optimization problem, $Q_3'(n, t_1, t_2)$, is converted to the corresponding crisp nonlinear optimization problem. The crisp nonlinear problem, $Q_3''(n, t_1, t_2)$, is given by,
\begin{equation*}
\begin{split}
Q_3''(n, t_1, t_2): \qquad &\underset{\left(n, t_{1}, t_{2}\right)} {\operatorname{Max}}\quad  \phi \\
\text { such that} \qquad &\phi - \left( \frac{\text{ETC}_{UB} - Z_L}{Z_U - Z_L}\right) \leq 0,\\
& \phi - \left( \frac{\alpha + b_1 - g_3(n, t_1, t_2)}{b_1}\right) \leq 0,\\
& \phi - \left( \frac{\beta + b_2 - h_3(n, t_1, t_2)}{b_2}\right) \leq 0,\\
\end{split}
\end{equation*}
 $\text{where} \ Z_{U} = \max \left\{Z_1(n, t_1, t_2), Z_2(n, t_1, t_2)\right\},\ 
    Z_{L} = \min \left\{Z_1(n, t_1, t_2), Z_2(n, t_1, t_2)\right\},$
$t_{1},\ t_{2}>0$,\ $t_{2}>t_{1},\ t_2 \leq \lambda_0$, \ $\textrm{\textit{a}} > 0$, $\textrm{\textit{a}} > \lambda_j, j=0,1$, $b_i \in [0, 1], i = 1, 2$ and $ \phi \in [0, 1]$.

The crisp optimization problem, $Q_3''(n, t_1, t_2)$ is solved by the fmincon solver in MATLAB, and examples for various values of fuzzy and crisp parameters are illustrated in  Table~\ref{tab:max} and Table~\ref{tab:maxcrisp}, respectively to verify the computation and also for a comparative study.
\subsection{Acceptance sampling plan under Type I censoring}
In this ASP, from the given lot, a random sample of size $n$ is taken and tested up to a predetermined time. Acceptance and rejection of the lot are decided upon the fuzzy hypothesis given in~(\ref{2.3}). Our ASP, in this case, is as follows: 
\begin{enumerate}
    \item Take a random sample of size $n$ and run the test up to time $\tau$. Let $q$ be the number of failures obtained with corresponding lifetimes $X_{1, n}, X_{2, n}, \ldots, X_{q, n}$.
\item  The maximum likelihood estimate of $\lambda$ is given by,
\scalebox{1.2}{$
\hat{\lambda}=\frac{\sum\limits_{i=i}^{q} X_{i, n}+(n-q) \tau}{q}.
$}
\item If $t_{1} \leq \hat{\lambda}<t_{2}$, repeat Steps 1 and 2; else go to Step 4.
\item If $\hat{\lambda}<t_{1}$, reject the lot. If $\hat{\lambda} \geq t_{2}$, accept the lot.
\end{enumerate}
But $\hat{\lambda}$ is asymptotically normal with mean $\lambda$ and standard deviation
$
\frac{\lambda}{\sqrt{n\left(1-\exp \left(\frac{-\tau}{\lambda}\right)\right)}}
$
(see \citeN{bartholomew1963sampling}).
By taking all the parameters as fuzzy numbers, we get the required probabilities for the acceptance sampling plan after simplification under $H_j, j=0,1$ as follows:
\begin{equation}
p_{c}=p\left(t_{1} \leq \hat{\lambda}<t_{2}\right)
=p\left(\frac{t_{1}-\lambda_j}{\lambda_j} \sqrt{n\left(1-\exp \left(\frac{-\tau}{\lambda_j}\right)\right)} \leq Z<\frac{t_{2}-\lambda_j}{\lambda_j} \sqrt{\left.n\left(1-\exp \left(\frac{-\tau}{\lambda_j}\right)\right)\right)}\right.,
 \label{3.21}
 \end{equation}
 \vspace{-2mm}
 \begin{equation}
   p_{a}=p\left(t_{2} \leq \hat{\lambda}\right)=p\left(Z \geq \frac{t_{2}-\lambda_j}{\lambda_j} \sqrt{n\left(1-\exp \left(\frac{-\tau}{\lambda_j}\right)\right)}\right), 
   \label{3.22}  
 \end{equation}
 \begin{equation}
p_{r}=p\left(\hat{\lambda}<t_{1}\right)=p\left(Z<\frac{t_{1}-\lambda_j}{\lambda_j} \sqrt{n\left(1-\exp \left(\frac{-\tau}{\lambda_j}\right)\right)}\right), 
\label{3.23}
\end{equation}
where $Z$ is the standard normal random variable. Our problem is to find $\left(n, t_{1}, t_{2}\right)$ with minimum total testing cost. Since total testing cost is a random quantity, we minimize the expected testing cost at $\lambda_{0}$ subject to fuzzy hypothesis given by~(\ref{2.3}) and obtain the optimal values of $(n, t_1, t_2)$. Since there is a truncation in total testing time $\tau$,
two alternatives exist such as the testing can be terminated at $\tau$ or  testing can be stopped, if all the test items fail before time $\tau$. Assume that $\tau_0$ represents the time when every test item for a specific sample has failed. Then, $\tau_{0} \leq \tau$. The total number of samples that need to be tested is $\frac{1}{1-p_{c}}$. Because of this, the overall testing time is $\frac{\tau}{1-p_{c}}$ or less. The total testing expense will be less than or equal to $\frac{C \tau}{1-p_{c}}$, which is a random amount, given that $C$ represents the cost of testing a sample for a unit of time. However, the actual cost can be less. Since it is difficult to minimize the random cost, we consider minimizing the expected value of the upper bound for random testing cost, which is denoted by $\mathrm{ETC}_{\mathrm{ub}}$ and is given by 
$$\mathrm{ETC}_{\mathrm{ub}}=\frac{C \times \tau}{1-p_{c}}.$$

\subsubsection{Fuzzy optimization problem and solution procedure}

The optimal strategy for minimising $\mathrm{ETC}_\mathrm{ub}$ will depend on $\lambda$ because $\mathrm{ETC}_\mathrm{ub}$ is a function of $\lambda$. Since there is no prior distribution assumed, therefore $\lambda$ is unknown. To get beyond this issue, we take $\mathrm{ETC}_{\mathrm{ub}}$ under $H_0$. Our basic objective is to reduce $\mathrm{ETC}_{\mathrm{ub}}$ subject to fuzzy Type I and Type II errors based on the fuzzy hypothesis given by~(\ref{2.3}). Since $\lambda_{0}$ is the AQL, we have $t_{2} \leq \lambda_{0}$ (applying the ASP definition provided in Section 3.3). Thus we have the fuzzy optimization problem $P_4(n, t_1, t_2)$:
\begin{equation*} 
\begin{split}
P_4(n, t_1, t_2): \qquad &\underset{\left(n,t_{1}, t_{2}\right)}{\operatorname{Min}}\left(\frac{C \times \tau}{1-p_{c}}\right) \\
\text {such that} \qquad &P\left( \text{Type I error}  \mid H_0 \right) \stackrel{\sim}{\leq} \alpha, \\
&P\left(\text {Type II error} \mid H_1\right) \stackrel{\sim}{\leq} \beta,
\end{split}
\end{equation*}
where $\alpha$, $\beta$ are fuzzy and $t_{1}, t_{2}>0$ and $t_{2}>t_{1}$. That is,
\begin{equation*}
\begin{split}
&\underset{\left(n,t_{1}, t_{2}\right)}{\operatorname{Min}}\left(\frac{C \times \tau}{1-p_{c}}\right) \text{under} \ H_0 \\
\text { such that} \qquad &\left(\frac{p_{r}}{1-p_{c}} \mid H_0 \right) \stackrel{\sim}{\leq} \alpha, \\
&\left(\frac{p_{a}}{1-p_{c}} \mid H_1 \right) \stackrel{\sim}{\leq} \beta,\qquad \qquad \qquad
\end{split}
\end{equation*}
where $t_{1}, t_{2}>0$ and $t_{2}>t_{1}$. Substituting for $p_{c}, p_{a}$ and $p_{r}$ from~(\ref{3.21}),~(\ref{3.22}) and~(\ref{3.23}), the fuzzy optimization problem $P_4(n, t_1, t_2)$ becomes $P_4'(n, t_1, t_2)$, given by,
\begin{equation}
\begin{aligned}[b]
P_4'(n, t_1, t_2): \qquad &\underset{\left(n,t_{1}, t_{2}\right)}{\operatorname{Min}} \mathrm{ETC}_{\mathrm{ub}} \\
\text { such that} \qquad & g_4(n, t_1, t_2) \stackrel{\sim}{\leq} \alpha,\\
& h_4(n, t_1, t_2) \stackrel{\sim}{\leq} \beta,  
\end{aligned}
\label{3.24}
\end{equation}
 \begin{equation*}
    \begin{split}
    \textrm{where}\ \ \mathrm{ETC}_{\mathrm{ub}} &= \frac{C \times \tau}{1-p\left(\frac{t_{1}-\lambda_0}{\lambda_0} \sqrt{n\left(1-\exp \left(\frac{-\tau}{\lambda_0}\right)\right)} \leq z<\frac{t_{2}-\lambda_0}{\lambda_0} \sqrt{\left.n\left(1-\exp \left(\frac{-\tau}{\lambda_0}\right)\right)\right)}\right.},\qquad \qquad\\
    g_4(n, t_1, t_2) &= \frac{p\left(z<\frac{t_{1}-\lambda_0}{\lambda_0} \sqrt{n\left(1-\exp \left(\frac{-\tau}{\lambda_0}\right)\right)}\right)}{1-p\left(\frac{t_{1}-\lambda_0}{\lambda_0} \sqrt{n\left(1-\exp \left(\frac{-\tau}{\lambda_0}\right)\right)} \leq z<\frac{t_{2}-\lambda_0}{\lambda_0} \sqrt{\left.n\left(1-\exp \left(\frac{-\tau}{\lambda_0}\right)\right)\right)}\right.},\qquad \qquad \qquad \qquad\\
        h_4(n, t_1, t_2) &= \frac{p\left(z \geq \frac{t_{2}-\lambda_1}{\lambda_1} \sqrt{n\left(1-\exp \left(\frac{-\tau}{\lambda_1}\right)\right)}\right)}{1-p\left(\frac{t_{1}-\lambda_1}{\lambda_1} \sqrt{n\left(1-\exp \left(\frac{-\tau}{\lambda_1}\right)\right)} \leq z<\frac{t_{2}-\lambda_1}{\lambda_1} \sqrt{\left.n\left(1-\exp \left(\frac{-\tau}{\lambda_1}\right)\right)\right)}\right.},\qquad \qquad \qquad \qquad\\
       t_{1}, t_{2}>0 \text{ and } &t_{2}>t_{1}.
    \end{split}
\end{equation*}
The fuzzy nonlinear optimization problem, $P_4'(n, t_1, t_2)$, is solved by changing it to its crisp alternative $P_4''(n, t_1, t_2)$ by applying the steps in the algorithm for fuzzy optimization explained in the Subsection 2.1.1., by taking the membership functions (2.14 and 2.15) of $\alpha$ and $\beta$. The resulting crisp optimization problem $P_4''(n, t_1, t_2)$ is given by, 
\begin{equation*}
\begin{split}
P_4''(n, t_1, t_2): \qquad &\underset{\left(n, t_{1}, t_{2}\right)} {\operatorname{Max}}\quad  \phi \\
\text { such that} \qquad &\phi - \left( \frac{\mathrm{ETC}_{\mathrm{ub}} - Z_L}{Z_U - Z_L}\right) \leq 0,\\
& \phi - \left( \frac{\alpha + b_1 - g_4(n, t_1, t_2)}{b_1}\right) \leq 0,\\
& \phi - \left( \frac{\beta + b_2 - h_4(n, t_1, t_2)}{b_2}\right) \leq 0,\\
\end{split}
\end{equation*}
$\text{where } \ Z_{U} = \max \left\{Z_1(n, t_1, t_2), Z_2(n, t_1, t_2)\right\}, \
    Z_{L} = \min \left\{Z_1(n, t_1, t_2), Z_2(n, t_1, t_2)\right\},
t_{1}, \ t_{2}>0$, $t_{2}>t_{1}, \ b_i \in [0, 1],\ i = 1, 2$ and $ \phi \in [0, 1]$. 

The optimal values of required parameters $n$, $t_1$ and $t_2$ are found by solving $P_4''(n, t_1, t_2)$ using fmincon solver in MATLAB. In Table~\ref{tab:TypeI},  examples for various parameter values are tabulated. Also, a comparative study with corresponding crisp values of parameters is also recorded in Table~\ref{tab:TypeIcrisp}.

\section{De-fuzzification Methods and their limitations in Acceptance Sampling Plans}

De-fuzzification is choosing a meaningful crisp value in relation to a fuzzy set.  Many applications employ standard de-fuzzification methods like the center of gravity to generate crisp output. It is observed that most of the standard de-fuzzification methods give the same crisp output for the fuzzy numbers $H_j(\lambda), j=0,1, \alpha$ and $\beta$ as that of the method of center of gravity. Thus, for the purpose of applying de-fuzzification methods to the acceptance sampling plans discussed in previous sections, a widely used de-fuzzification method, namely center of gravity method for the fuzzy number $\lambda_j, j = 0, 1$ with membership function given by,
\begin{equation*}
    H_j(\lambda)=\left\{\begin{array}{ll}
\frac{1+cos(\textit{a}\pi(\lambda-\lambda_j))}{2}  & , \quad \lambda_j - \frac{1}{\textit{a}} \leq \lambda \leq \lambda_j + \frac{1}{\textit{a}}, \\
0  & ,\quad \text { otherwise, } \qquad \qquad \qquad j= 0, 1.
\end{array}\right.
\end{equation*}
In this method, the corresponding crisp value is computed as,
\begin{equation}
    \lambda_j^* = \frac{ \displaystyle\int_{\lambda_j - \frac{1}{a}}^{\lambda_j + \frac{1}{a}} H_j(\lambda) \cdot \lambda \ d\lambda }{\displaystyle\int_{\lambda_j - \frac{1}{a}}^{\lambda_j + \frac{1}{a}} H_j(\lambda)\ d\lambda }
    = \lambda_j. 
    \label{4.1}
\end{equation}
So this de-fuzzification method in designing acceptance sampling plans produces results similar to those discussed in \citeN{MK}. Hence, while designing acceptance sampling plans, the uncertainty in the data is not taken care of by de-fuzzification methods such as the center of gravity method. Therefore, it is evident from~(\ref{4.1}) that the acceptance sampling plan obtained using the weighted probability density function dealt with fuzzy hypotheses in a better way than any de-fuzzification methods.
\section{Numerical Results, Comparison and Case Study}
 In this section, the computational outcomes of all the above sections are discussed. The numerical results for various sampling plans with different parameter values in the case of fuzzy parameters are presented in Tables~\ref{tab:SSP}, \ref{tab:min}, \ref{tab:max}, and \ref{tab:TypeI} and their corresponding results in crisp case are tabulated in Tables~\ref{tab:SSPcrisp}, \ref{tab:mincrisp}, \ref{tab:maxcrisp}, and \ref{tab:TypeIcrisp}.  In all the cases, the expected testing cost for fuzzy parameters is compared with the crisp case. It is observed from the tables that, the cost of testing in the fuzzy case approaches that of the crisp case as the fuzzy parameter values come close to the crisp value, which validates the sampling plans. For example, in Table~\ref{tab:SSP} of sequential sampling plan, we have \lq\lq$\lambda_{0}$ is approximately 300", \lq\lq$\lambda_{1}$ is approximately 50", \lq\lq$\alpha$ is close to 0.05", \lq\lq$\beta$ is close to 0.05" for $a=1500$, our sampling plan (for $C=1$) gives expected testing cost of $665.7614$,  and for $a=15000$, ETC is reduced to $661.2965$ whereas in the crisp case, i.e, $\lambda_{0}=300, \lambda_{1}=50, \alpha=0.05$ and $\beta=0.05$, the corresponding ETC is 654.1617 which shows the convergence of ETC in fuzzy case to crisp as the fuzzy numbers approaches to exact value.  Table~\ref{tab:comparison}  illustrates the comparison of ETCs associated with all acceptance sampling plans with fuzzy parameters. Consider the following example (in Table~\ref{tab:comparison}) for the set of fuzzy parameters: $\lambda_{0}=500,\ \lambda_{1}=150,\ \alpha=0.05,\ \beta=0.05,\ a = 15000$, the ETC obtained for the acceptance sampling plans, namely, sequential sampling plan, RGSP based on minimum of $n$ observations, maximum $n$ observations, and RGSP based on Type I censoring are 3866.4515, 491341.2, 2003.4707 and 99.9722, respectively. These comparisons show that the RGSP with Type I censoring has the lowest testing cost of all the acceptance sampling plans discussed here for fuzzy parameters.
\subsection{Case study: A real example}
To investigate the different acceptance sampling plans employed in this work using fuzzy hypotheses, consider a real-world case study discussed in \citeN{alizadeh2015testing}. The data presented here come from an experiment of testing new compact electrical appliance models. The lifespans shown here are the number of cycles of use that have been completed before the appliances break down, the appliances were operated repeatedly by an automatic testing system. This data follows exponential distribution.
 
$$170, 2694, 3034, 3214, 2400, 2471, 329, 13403, 381, 708, 7846, 1062, 1594, 1925, 1990, 2223, 2327, 2451, 2551,$$
$$2565, 2568, 2702, 11, 35, 2761, 2831, 3059, 3112, 3478, 3504, 4329, 6367, 6976, 1167, 49, 958.$$

Let the plan parameters be given in linguistic variables as $\lq\lq\lambda_{0}$ is nearly $3000"$, $\lq \lq\lambda_{1}$ is close to $600"$, $\alpha$ and $\beta$ are $\lq \lq$approximately $0.05"$. These uncertainties are handled by taking them as fuzzy numbers with appropriate fuzzy number representations. The optimization problems $P_1''(t_1, t_2)$, $P_2''(n, t_1, t_2)$, $P_3''(n, t_1, t_2)$, and $P_4''(n, t_1, t_2)$ are solved for the plan parameters $t_{1}$ and $t_{2}$ with $a=18000$ and $b_1 = b_2 =0.05$. The various acceptance sampling plans discussed in this article produce the following results for the aforementioned data:
\begin{enumerate}
    \item The sequential sampling plan gives $t_{1}=41$ and $t_{2}=3159$. Based on the data set, we accept the lot, at the point of $8^{th}$ failure point, as $Y_i, i=1,2,\cdots 7$ is in between $t_1$ and $t_2$ and $Y_8$ is greater than $t_{1}=41$.
\item The RGSP results in $t_1 = 4$, $t_2 = 141$ and $n = 20$ based on minimum of $n$ observations. We then take into account, the first 20 observations from the preceding data. The minimum of these 20 observations in this instance exceeds 141. So we can accept the lot.
\item Based on the RGSP, using the maximum of $n$ observations, we obtain $t_1=203,\ t_2=2630$ and $n=2$. The first two observations from the above data are then taken into consideration and their maximum is 2694, which is higher than 2630. Therefore, we accept the lot.
\item Based on Type I censoring, the RGSP yields the values $t_1=1219$, $t_2=1990$ and $n=13$. We then take into account, the first 13 items using the data provided above. Find the average lifespan, when the number of cycles is truncated at $T=2000$. The 13 observations in the data have an average lifespan of $3040.6667$, which is higher than $t_2=1990$. Therefore, we accept the lot.
\end{enumerate}

%----------------------------------Table 1 Sequential sampling plan----------------
\begin{acmtable}{350pt}[!htb]
\centering
\begin{tabular}{llllllll|lll}
\hline
$\lambda_0$ & $\lambda_1$  &$\alpha$&$\beta$&$a$& \multicolumn{3}{c|}{$Q_1''(t_1, t_2)$} & \multicolumn{3}{c}{$P_1''(t_1, t_2)$}\\
\cline { 6- 11} &&&&& $t_1$ &$t_2$&$ETC_{UB} $ &$t_1$ &  $t_2$&$ETC^{*}$\\
\hline
300&50&0.05&0.05&
1500&7.0971&247.9307&1317.112&5.8231&251.1178&665.7614\\
&&&&15000&6.2579&254.3787&1336.622&6.4907&251.617&661.2965\\
&&0.05&0.1&1500&8.271&205.0141&1141.6933&8.1902&205.6224&567.5001\\
&&&&15000&8.2609&205.0045&1127.7786&7.6285&204.1645&564.5195\\
 300 & 70 &0.1&0.1 &2100&12.4249&268.0108&1341.6862&10.8498& 289.1994&720.7401\\
&&&&15000&13.6293&274.7578&1349.7204&13.5624& 275.2881&676.2045\\
 &  &0.05& 0.05&2100&4.2697&393.4824&2126.2055&3.8761&412.3522&1128.9831\\
 &&&&15000&4.0084&406.8951&2215.1938&3.9803&407.1574&1107.8074\\
 \hline
 \end{tabular}
\caption{Sequential sampling plan with fuzzy parameters for C=1 and $b_1 = b_2 = 0.05$.}
\label{tab:SSP}
\end{acmtable}

\begin{acmtable}{230pt}
\centering
\begin{tabular}{llllllllll}
\hline
$\lambda_0$&$\lambda_1$&$\alpha$&$\beta$&$t_1$&$t_2$&$ETC$\\
\hline
300&50&0.05&0.05&6.9591&275.6183&654.1617\\
&&0.05&0.1&8.0812&204.9714&564.3796\\
300&70&0.1&0.1&13.6317&275.0377&675.3388\\
&&0.05&0.05&4.0960&406.8418&1106.1365\\
 \hline
\end{tabular}
\caption{Sequential sampling plan with crisp parameters for C=1.} 
\label{tab:SSPcrisp}
\end{acmtable}
%--------------------Table 2 Minimum----------------
\begin{acmtable}{400pt}
\centering
\scriptsize{
\begin{tabular}{lllllllll|llll}
\hline
$\lambda_0$ & $\lambda_1$  &$\alpha$&$\beta$&$a$& \multicolumn{4}{c|}{$Q_2''(n, t_1, t_2)$ } & \multicolumn{4}{c}{$P_2''(n, t_1, t_2)$}\\
\cline { 6- 13} &&&&& $t_1$ &$t_2$&$n$&$ETC_{UB} $ &$t_1$ &  $t_2$&$n$&$ETC^{*}$\\
\hline 300&200&0.05&0.05&1500&0.00001&81.5216& 42& 863623.0499&0.000001& 79.3124& 50 &202277.1689\\
&&&&15000&0.0000037&70.1102&48&888411.9581&0.0000041&65.3243&50&308804.9568\\
&&0.05&0.1&1500&0.0005&283.8829&10&531916.8655&0.0008&283.9992&9&111118.3155\\
&&&&15000&0.0004&320.7604&9&850862.4562&0.0005&284.0415&9&155357.7843\\
500&300&0.05&0.05&1500&0.00011&158.5504&30&240876.2939&0.0006&177.8714&23&46681.3855\\
&&&&15000&0.0004&137.4607&29&93348.8731&0.0004&134.4211&30&49062.1925\\
&&0.05&0.1&1500&0.0002&203.9995&17&44642.5866&0.0027&204.2658&17&20676.9203\\
&&&&15000&0.0021&203.7636&17&55787.4515&0.0018&224.5448&15&26779.9451\\
 \hline
\end{tabular}}
\caption{RGSP by minimum of $n$ observations with fuzzy parameters for C=1 and $b_1$ = $b_2$ = $0.05$.}
\label{tab:min}
\end{acmtable}

\begin{acmtable}{260pt}
\begin{tabular}{llllllllll}
\hline
$\lambda_0$&$\lambda_1$&$\alpha$&$\beta$&$t_1$&$t_2$&$n$&$ETC$\\
\hline
300&200&0.05&0.05&0.00001& 65.801& 50 &316930.5172\\
&&0.05&0.1&0.00012&284.172&10&370615.1316\\
500&300&0.05&0.05&0.0003&134.4514&30&50246.7326\\
&&0.05&0.1&0.0015&204.3024&17&29028.4419\\
 \hline
\end{tabular}
\caption{RGSP by minimum of $n$ observations for crisp parameters for C=1.}
\label{tab:mincrisp}
\end{acmtable}
%------------Table 3 MAximum---------------
\begin{acmtable}{400pt}
\centering
{\scriptsize
\begin{tabular}{lllllllll|llll}
\hline 
 $\lambda_0$ & $\lambda_1$  &$\alpha$&$\beta$&$a$& \multicolumn{4}{c|}{$Q_3''(n, t_1, t_2)$ } & \multicolumn{4}{c}{$P_3''(n, t_1, t_2)$}\\
\cline { 6- 13} &&&&& $t_1$ &$t_2$&$n$&$ETC_{UB} $ &$t_1$ &  $t_2$&$n$&$ETC^{*}$\\
\hline 300&50&0.05&0.05&1500&141.864&326.2226& 9& 1703.2061&130.9470&338.7602&11&921.1108\\
&&&&15000&135.3401&332.8650&9&1678.8201&130.6584&338.9876&12&916.1062\\
&&0.05&0.1&1500&84.0375&290.9444&5&1510.1843&176.205&197.1019&5&634.73\\
&&&&15000&51.922&322.6559&3&1287.7343&176.3513&196.9506&5&631.3610\\
500&150&0.05&0.05&1500&285.8609&887.9414&4&3811.4313&155.5672&997.4723&4&2039.2569\\
&&&&15000&310.9016&864.9856&6&3437.3812&224.4646&938.2128&4&2003.4707\\
&&0.05&0.1&1500&200.0049&621.0167&3&2290.9583&199.9814&621.0203&3&1366.8428\\
&&&&15000&197.0036&621.0157&3&2223.5466&196.998&621.0034&3&1111.5077\\
 \hline
\end{tabular}}
\caption{RGSP by maximum of $n$ observations with fuzzy parameters for C=1 and $b_1 = b_2 = 0.05$.}
\label{tab:max}
\end{acmtable}

\begin{acmtable}{260pt}
\begin{tabular}{llllllllll}
\hline
$\lambda_0$&$\lambda_1$&$\alpha$&$\beta$&$t_1$&$t_2$&$n$&$ETC$\\
\hline
300&50&0.05&0.05&239.5461& 228.7969& 5 &620.609\\
&&0.05&0.1&192.6274&181.4727&4&545.6562\\
500&150&0.05&0.05&441.8257&756.6208&6&1399.2133\\
&&0.05&0.1&200.7733&621.8946&3&1110.014\\
 \hline
\end{tabular}
\caption{RGSP by maximum of $n$ observations with crisp parameters for C=1.}
\label{tab:maxcrisp}
\end{acmtable}
%--------------------Table 4 Type 1---------
\begin{acmtable}{270pt}
\begin{tabular}{llllllllll}
\hline
$\lambda_0$&$\lambda_1$&$T$&$\alpha$&$\beta$&$t_1$&$t_2$&$n$&$ETC_{ub}$\\
\hline
300&200&50&0.01&0.01&236.8898&236.8898&33&49.9885\\
&&&0.05&0.05&245.5488&245.5488&30&49.9887\\
300&200&100&0.01&0.01&235.182&235.182&28&99.9730\\
&&&0.01&0.05&227.6252&227.6252&46&99.9722\\
 \hline
\end{tabular}
\caption{RGSP with Type I censoring with fuzzy parameters for C=1 and  $b_1 = b_2 = 0.01.$}
\label{tab:TypeI}
\end{acmtable}

\begin{acmtable}{270pt}
\begin{tabular}{llllllllll}
\hline
$\lambda_0$&$\lambda_1$&$T$&$\alpha$&$\beta$&$t_1$&$t_2$&$n$&$ETC_{ub}$\\
\hline
300&200&50&0.01&0.01&237.1385&237.1385&33&50\\
&&&0.05&0.05&246.520&246.5220&30&50\\
300&200&100&0.01&0.01&234.6239&234.6239&28&100\\
&&&0.01&0.05&227.449&227.449&46&100\\
 \hline
\end{tabular}
\caption{RGSP with Type I censoring with crisp parameters for C=1.}
\label{tab:TypeIcrisp}
\end{acmtable}
%--------------------Table 6 Comparison-------
\begin{acmtable}{300pt}
\begin{tabular}{llllllll}
\hline
$\lambda_0$&$\lambda_1$&$\alpha$&$\beta$&$A$&$B$&$C$&$D$\\
\hline
300&50&0.05&0.05&661.2965&$4.065681 \cdot 10^4$&916.1062&49.8877\\
300&200&0.05&0.05&1.8562$\cdot 10^4$&$3.088049 \cdot 10^5$&4606.3980&49.9887\\
500&150&0.05&0.05&3866.4515&$4.913412 \cdot 10^5$&2003.4707&99.9722\\
 \hline
\end{tabular}
\caption{Comparison of expected testing costs in the sequential sampling plan (column A), RGSP based on minimum of $n$ observations (column B), maximum of $n$ observations (column C) and RGSP under Type I censoring (column D) for fuzzy parameters with C=1, $a = 15000$ and  $b_1 = b_2 = 0.05.$}
\label{tab:comparison}
\end{acmtable}

\section{Conclusion}
The majority of acceptance sampling plans available in the existing literature are based on the assumption that data are crisp and assume that there is no uncertainty in the data. Thus, no sampling plans in the literature address the need of including ambiguous parameter information in the design of acceptance sampling plans. In this study, we designed acceptance sampling plans by incorporating uncertainties in data and solving a series of fuzzy optimization problems to obtain the total expected cost of testing when all parameters in the sampling plan are expressed in linguistic variables. Note that, all the parameters AQL, RQL, $\alpha$, and $\beta$ are taken as fuzzy numbers. Observe that, the objective function considered in optimization problems is the ETC and which is minimized, satisfying the constraints of Type I and Type II errors. This was necessary, due to the reason that the actual costs involved in all the sampling plans are random in nature and thereby the corresponding optimization problems are intractable. However, in practice, the actual cost involved in testing may be less than the ETC reported here. For example, in RGSP based on minimum of $n$ observations for fuzzy parameters (in Table~\ref{tab:min}), when \lq\lq$\lambda_{0}$ is approximately 300", \lq\lq$\lambda_{1}$ is approximately 200", \lq\lq$\alpha$ is close to 0.05", \lq\lq$\beta$ is close to 0.05" for $a=1500$ and  $C=1$, we get  ETC = $202277.1689$ and for $a=15000$, ETC becomes $308804.9568$, but in the crisp case, i.e, $\lambda_{0}=300, \lambda_{1}=200, \alpha=0.05$ and $\beta=0.05$, the corresponding ETC is 316930.5172. Next, consider another example in Table~\ref{tab:SSP} of sequential sampling plan for fuzzy parameters, when \lq\lq$\lambda_{0}$ is approximately 300",\lq\lq$\lambda_{1}$ is approximately 50", \lq\lq$\alpha$ is close to 0.05", \lq\lq$\beta$ is close to 0.05" for $a=1500$, our ASP (for $C=1$) gives ETC of $665.7614$ and for $a=15000$, ETC is reduced to $661.2965$  whereas in the crisp case, i.e, $\lambda_{0}=300, \lambda_{1}=50, \alpha=0.05$ and $\beta=0.05$, the corresponding ETC is 654.1617 which shows the convergence of ETC in fuzzy case to crisp as the fuzzy numbers approaches to exact value. We observe from our computational experience that, some of the fuzzy acceptance sampling plans have reported slightly bigger costs than that in the existing ASP (see \citeN{MK}). On the other hand, the corresponding cost is lower than that of crisp ASP. The lower cost in crisp ASP may be due to the assumption that the parameters are crisp, which is not true in practice, it can be vague, which is to be taken into account while designing ASP. So in this work, we have solved fuzzy optimization problems to design acceptance sampling plans based upon uncertainties in the parameters which is more practical and real. In Section 4, we discussed, de-fuzzification methods and their limitations in ASP, it is noted that certain de-fuzzification methods such as the center of gravity method, do not handle the fuzziness in the parameters, thereby the results obtained exactly coincide with those results obtained with crisp case (see \citeN{MK}). Hence, we have used the weighted probability density function to incorporate the uncertainties in the parameter effectively. Table~\ref{tab:comparison} compares the testing costs for the acceptance sampling plans with fuzzy parameters. For the set of parameters $\lambda_{0}=500, \lambda_{1}=150, \alpha=0.05, \beta=0.05, a = 15000$, the expected cost obtained for the acceptance sampling plans, namely, sequential sampling plan, RGSP using minimum of $n$ observations, maximum $n$ observations and RGSP based on Type I censoring are 3866.4515, 491341.2, 2003.4707, and 99.9722, respectively. These comparisons demonstrate that the RGSP based on Type I censoring has the lowest testing cost. To illustrate the actual use of our acceptance sampling plans, a real-world example is also presented. The results obtained in this work can also be extended for more general class of lifetime distributions such as Weibull distribution based upon any other censoring schemes such as hybrid censoring.  
\bibliographystyle{acmtrans}
\bibliography{references}
\end{document}